\documentstyle[aps,prd,amssymb,amsmath,epsfig]{revtex}

\newcommand{\rmd}{{\rm d}}
\newcommand{\rme}{{\rm e}}
\newcommand{\mbf}{{\mathbf{f}}}
\newcommand{\bg}{{\mathbf{g}}}
\newcommand{\bk}{{\mathbf{k}}}
\newcommand{\bq}{{\mathbf{q}}}

\newcommand{\bu}{{\mathbf{u}}}
\newcommand{\bv}{{\mathbf{v}}}
\newcommand{\bx}{{\mathbf{x}}}
\newcommand{\bS}{{\mathbf{S}}}
\newcommand{\bU}{{\mathbf{U}}}
\newcommand{\bX}{{\mathbf{X}}}
\newcommand{\BE}{{\mathbb{E}}}
\newcommand{\BR}{{\mathbb{R}}}

\newcommand{\CA}{{\cal A}}
\newcommand{\CB}{{\cal B}}
\newcommand{\CC}{{\cal C}}
\newcommand{\CD}{{\cal D}}

\newcommand{\CJ}{{\cal J}}
\newcommand{\CR}{{\cal R}}
\newcommand{\CU}{{\cal U}}
\newcommand{\average}[1]{\left\langle #1 \right\rangle_\CD}
\newcommand{\laverage}[1]{\left\langle #1 \right\rangle_{\CD_{\rm \bf i}}}
\newcommand{\baverage}[1]{\left\langle #1 \right\rangle_{\CB_R}}
\newcommand{\initial}[1]{{#1_{\rm \bf i}}}
\newcommand{\inI}{{\bf I}}
\newcommand{\inII}{{\bf II}}
\newcommand{\inIII}{{\bf III}}
\begin{document}
\draft

\title{Backreaction of inhomogeneities on the expansion:\\
the evolution of cosmological parameters}
\author{Thomas Buchert\thanks{email: buchert@theorie.physik.uni-muenchen.de},}
\address{Theoretical Astrophysics Division, National Astronomical
Observatory, 2--21--1 Osawa Mitaka Tokyo 181--8588, Japan}
\author{Martin Kerscher\thanks{email: kerscher@theorie.physik.uni-muenchen.de},
and Christian Sicka\thanks{email: sicka@theorie.physik.uni-muenchen.de}}
\address{Theoretische Physik, Ludwig--Maximilians--Universit\"{a}t,
Theresienstra{\ss}e 37, 80333 M\"{u}nchen, Germany}
\date{submitted December 17, 1999, accepted May 3, 2000}
\maketitle
%
\pacs{98.80.-k, 98.80.Es, 95.35.+d, 98.80.Hw, 04.20.Cv}

\begin{abstract}
  Averaging   and    evolving   inhomogeneities   are   non--commuting
  operations.  This implies the existence of deviations of an averaged
  model from the  standard Fried\-mann--Lema\^\i{}tre cosmologies.  We
  quantify these  deviations, encoded in a  backreaction parameter, in
  the framework  of Newtonian cosmology.  We employ  the linear theory
  of  gravitational   instability  in  the   Eulerian  and  Lagrangian
  approaches,  as  well  as  the  spherically--  and  plane--symmetric
  solutions as  standards of  reference.  We propose  a model  for the
  evolution  of the average  characteristics of  a spatial  domain for
  generic  initial  conditions that  contains  the spherical  top--hat
  model and the  planar collapse model as exact  sub cases.  A central
  result  is   that  the  backreaction  term   itself,  calculated  on
  sufficiently large  domains, is small  but, still, its  presence can
  drive the  cosmological parameters on the averaging  domain far away
  from their  global values  of the standard  model.  We  quantify the
  variations of these  parameters in terms of the  fluctuations in the
  initial data as derived from the power spectrum of initial cold dark
  matter density  fluctuations.  E.g.\  in a domain  with a  radius of
  100Mpc today  and initially one--$\sigma$  fluctuations, the density
  parameters   deviate  from   their  homogeneous   values   by  15\%;
  three--$\sigma$ fluctuations lead to deviations larger than 100\%.
\end{abstract}

\twocolumn
\narrowtext

\section{Introduction}

The standard picture of the  evolution of the Universe relates typical
observables to global scalar  cosmological parameters such as the mean
matter density  and the global  expansion rate.  These  parameters are
derived   from   solutions  of   Einstein's   or   Newton's  laws   of
gravitationally interacting {\em  homogeneous} systems.  Moreover, the
class  of homogeneous  systems is  usually narrowed  to  the isotropic
ones,  in most  cases to  the  simplest, spatially  flat model,  first
discussed by Einstein and de Sitter  at the beginning of this century. 
This  {\em standard  model} has  survived in  spite of  the tremendous
development of  our knowledge about  inhomogeneities in the  Universe. 
The key argument of applicability of the homogeneous--isotropic models
is the conjecture that {\em  on average} the Universe may still follow
the evolution  of these simple models.  Without  specifying the notion
of averaging  and without entering  a sophisticated discussion  of how
one should  average an inhomogeneous  spacetime, it can  be definitely
said  that   this  conjecture  is  in  itself   courageous  given  the
nonlinearity   of  Einstein's  or   Newton's  laws,   the  long--range
attractive  forces,  and the  fact  that  the  solutions are  linearly
unstable  to  inhomogeneous perturbations  in  the  matter variables.  
Because  of  the  continued  success  of  the  standard  model  as  an
explanation  of a  variety  of orthogonal  observations,  the need  to
replace  the homogeneous--isotropic  models by  averaged inhomogeneous
models  is  not obvious  to  most  researchers  in cosmology  (from  a
physical point of view it should be obvious).  Notwithstanding we will
add a substantial contribution to  supporting this need in the present
article, shedding new light  on the interpretation of observations and
N--body simulations in  spatial domains that cover only  a few percent
of the Hubble volume.

Recent progress in the  understanding of the effective (i.e. spatially
averaged)  dynamical properties  of inhomogeneous  cosmological models
may be  summarized by  saying that one  can unambiguously  replace the
standard  Friedmann  equations  governing  the  homogeneous--isotropic
models   by  corresponding  equations   for  the   spatially  averaged
variables, provided  we focus our attention to  {\em scalar} dynamical
variables.  The  averaged equations  feature an additional  source, we
call  it  ``backreaction''  of  the  inhomogeneities  on  the  average
expansion,  which  will  be  the  basis of  our  investigations.   The
restriction  to scalars  has  its  root in  the  averaging problem  of
general     relativity     {}\cite{ellis:relativistic},    where     a
straightforward answer  to the question  of how one should  average an
inhomogeneous spacetime  metric is not  at hand.  For  scalars spatial
averaging can  be properly  defined, provided we  have a  foliation of
spacetime. The  results of averaging  the Newtonian equations  for the
evolution of a single dust component {}\cite{buchert:averaging}, where
also the choice  of foliation is not a problem, form  the basis of the
current  work.   The  corresponding  equations in  general  relativity
{}\cite{buchert:onaverage} will also be put into perspective.
 
We  shall   not  review   or  relate  other   numerous  work   on  the
``backreaction problem'',  partly because we want to  concentrate on a
quantitative  investigation  of  the generalized  Friedmann  equations
{}\cite{buchert:averaging},   and  partly   because   the  notion   of
``backreaction''  as we  employ  it differs  from  other notions;  for
example, many  authors study  the averaging problem  perturbatively in
general relativity  and consider as ``backreaction''  any deviation in
curvature   from   the   {\em   flat}  Friedmann   cosmologies.    The
approximations employed sometimes even imply that the perturbed models
remain within  the class of  Friedmann--Lema\^\i{}tre cosmologies.  To
review the  different averaging  procedures was recently  attempted by
Stoeger  et  al.~{}\cite{stoeger:averaging};   to  review  the  vastly
differing methods  of how the  ``backreaction'' can be estimated  is a
considerable,  but  desirable program  which  is  also complicated  by
different gauge choices in  a relativistic treatment.  We will mention
some of  these works  in the  main text mainly  to emphasize  that the
``backreaction'' we  are talking about  is not a  genuinely relativistic
effect, but is naturally present in the Newtonian framework as well.

Also, in the present work, we shift the emphasis to {\em fluctuations}
of the matter variables as a  result of backreaction, and we carry out
this investigation within a standard cosmological model in which there
is no {\em global}  contribution to backreaction.  In earlier, general
relativistic investigations researchers  mainly attempted to calculate
this  global   effect.   We  shall  quantify  the   influence  of  the
backreaction on the expansion in finite domains, small compared to the
horizon, using the Newtonian  approximation.  Consistently, we have to
assume that on some (very  large) scale the Universe expands according
to the Friedmann equation.   Possible boundary conditions that respect
the cosmological  principle are rare and practically  restricted to be
periodic in which case the boundary is empty and thus the backreaction
vanishes    globally    on     the    periodicity    scale    (compare
{}\cite{buchert:averaging},  especially Appendix  A).  We  will assume
that this scale is of the  order of the Hubble radius. Hence, with our
Newtonian treatment, we  are not able to give  any quantitative global
results, but we shall see that even domains with a size of hundreds of
Mpc's are influenced by backreaction.

Both  N--body  simulations and  most  analytical approximations  using
perturbation theory  rely on  the Newtonian approximation  to describe
the formation of structure in the Universe. The initial conditions are
often  specified  using  a  Gaussian  random  field  for  the  density
contrast.  These  two assumptions also  enter our calculations  of the
backreaction.  Note that both  the numerical and analytical approaches
{\em enforce}  a globally vanishing backreaction  by imposing periodic
boundary  conditions  on the  inhomogeneities.   However, contrary  to
current  N--body simulations,  we  are  not limited  by  box size  and
resolution,  but  we  are  limited  by the  validity  of  Zel'dovich's
approximation,  which we employ  as evolution  model.  The  latter has
been tested  thoroughly in comparison with N--body  simulations and we
shall add another test concerning its average performance.

We view the  present work as a dynamical  approach to cosmic variance:
not  only   the  density  contrast,  but  also   fluctuations  in  the
peculiar--velocity gradient (e.g. fluctuations in shear, expansion and
vorticity)  are  quantified by  their  effective  impact on  arbitrary
Newtonian portions of the Universe.  On such portions the interplay of
backreaction (condensed into an additional ``cosmological parameter'')
with  the  standard  parameters  of the  ``cosmic  triangle''  (matter
density parameter, ``curvature''  parameter and cosmological constant,
{}\cite{bahcall:triangle})    implies    scale--dependence   of    the
fluctuations of cosmological parameters.

This work may be approached with the following guidelines:
\begin{itemize}
\item  The ``cosmic  triangle'' fluctuates  on  a given  spatial scale.  
  Comparing observations on a given scale with cosmological parameters
  of  the   standard  model  has   to  be  subjected   to  statistical
  uncertainties of the parameters due to backreaction.
\item The  influence of backreaction may be  two--fold: for dominating
  shear fluctuations  the effective  density causing the  expansion is
  larger   than  the   actual   matter  density,   thus  mimicking   a
  ``kinematical dark  matter''; for dominating  expansion fluctuations
  backreaction  acts  accelerating, thus  mimicking  the  effect of  a
  (positive) cosmological constant.
\item Spatial  portions of  the Universe may  collapse as a  result of
  backreaction,  even   without  the  presence   of  over--densities.  
  Averaged   generic   inhomogeneities   invoke  deviations   in   the
  time--dependence  of  the  scale  factor  from  spherical  or  plane
  collapse.
\item Although in  some cases the backreaction source  term itself may
  be numerically small, its impact can be large, since already a small
  source term is  capable of driving the average  system far away from
  the homogeneous solutions.
\end{itemize}

We  have chosen to  organize the  investigation of  the effect  in the
following way.  In Sect.~\ref{sect:average-dynamics} we briefly review
the   generalized  Friedmann   equations   following  from   averaging
inhomogeneous dust cosmologies in  the Newtonian framework and provide
different representations  of the backreaction  that are used  in this
article.   We then move  in Sect.~\ref{sect:quantifying}  to dynamical
models  that  allow estimating  the  backreaction  term: the  Eulerian
linear  perturbation theory,  and the  Lagrangian  linear perturbation
theory (restricted to Zel'dovich`s approximation) serve as approximate
models.  The  plane and  spherical collapse models  serve as  exact
reference   solutions.    

We then illustrate  the results in Sect.~\ref{sect:evolution-cosmo} in
terms of the time evolution of cosmological parameters: we examine the
evolution of the scale factor  for typical initial conditions; then we
explore derived parameters: the Hubble--parameter and the deceleration
parameter as well as the cosmological density parameters.
In  Sect.~\ref{sect:some-remarks} we discuss  some related  issues: we
put  the  general  relativistic  equations into  perspective,  and  we
explain  the  role of  current  N--body  simulations  in view  of  our
results.

Technicalities         are         left         to         appendices:
Appendix~\ref{sect:generalized-friedmann} gives  a short derivation of
the  generalized expansion  law,  Appendix~\ref{sect:invariants} lists
useful writings of the principal  scalar invariants of a tensor field,
and  in  Appendix~\ref{sect:initial}  we  relate the  fluctuations  of
spatial invariants,  entering the ``backreaction` term,  to the initial
power spectrum of the inhomogeneities in the density contrast.

\section{Averaged dynamics in the Newtonian approximation}
\label{sect:average-dynamics}

In this section  we recall results obtained previously  by Buchert and
Ehlers  {}\cite{buchert:averaging} and summarize  alternative writings
of the backreaction source term.

\subsection{The generalized Friedmann equation}

In the Newtonian approximation the expansion of a domain is influenced
by the inhomogeneities inside the domain {}\cite{buchert:averaging}.

\begin{figure}
\begin{center}
\epsfig{figure=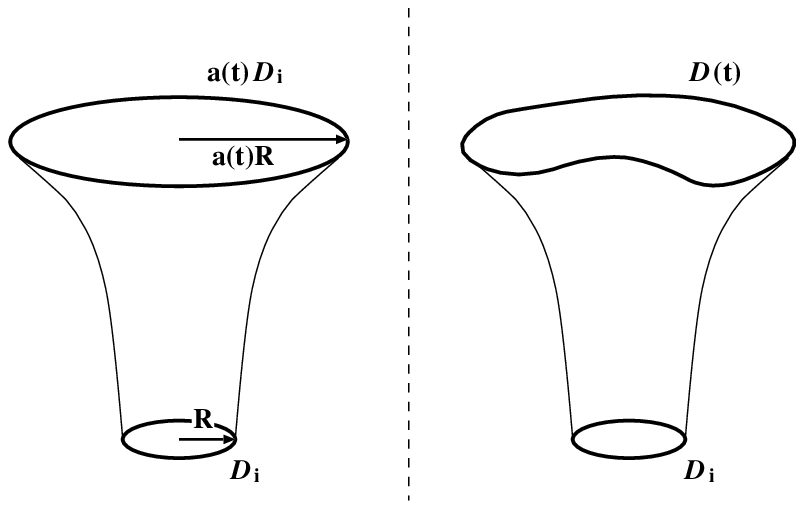,width=8cm}
\end{center}
\caption{\label{fig:domains}  The  evolution   of  an  initial  domain
$\initial{\CD}$ within the Hubble flow with global scale factor $a(t)$
(left) and for a general inhomogeneous setting (right).}
\end{figure}

We shall investigate fields  averaged over a simply--connected spatial
domain  $\CD$ at  time $t$  which evolved  out of  the  initial domain
$\initial{\CD}$  at time  $\initial{t}$, conserving  the mass  inside. 
The  averaged scale factor  $a_\CD$, depending  on content,  shape and
position of the spatial domain  of averaging $\CD$, is defined via the
domain's    volume    $V(t)=|\CD|$     and    the    initial    volume
$\initial{V}=V(\initial{t})=|\initial{\CD}|$:
\begin{equation}
\label{eq:def-ad}
a_\CD(t) = \left(\frac{V(t)}{\initial{V}}\right)^{\frac{1}{3}} .
\end{equation}
In Fig.~\ref{fig:domains}  we illustrate  the evolution of  an initial
domain  $\initial{\CD}$ with the  Hubble flow,  resulting in  a simple
rescaling  by the  global scale  factor  $a(t)$. In  a more  general
inhomogeneous  setting  the scale  factor  $a_\CD(t)$  is defined  via
Eq.~\eqref{eq:def-ad}.

With $\average{\cdot}$ we denote  spatial averaging in Eulerian space,
e.g.,  for a  spatial tensor  field  $\CA(\bx,t)=\{A_{ij}(\bx,t)\}$ we
simply have the Euclidean volume  integral normalized by the volume of
the domain:
\begin{equation}
\label{eq:average-def}
\average{\CA} = \frac{1}{V(t)} \int_\CD \rmd^3 x \; \CA(\bx,t) .
\end{equation}
For  domains  $\CD$ with  constant  mass  $M_\CD$,  as for  Lagrangian
defined domains, the average density is
\begin{equation}
\label{eq:average-rho}
\average{\varrho} = \frac{\laverage{\varrho(\initial{t})}}{a_\CD^3} 
= \frac{M_\CD}{a_\CD^3 \initial{V}} .
\end{equation}

Averaging Raychaudhuri's equation one  finds that the evolution of the
scale    factor    $a_\CD$    obeys    the    generalized    expansion
law\footnote{$\rmd_t$ is the  total (convective) time derivative, also
  denoted    with    an   over--dot    ``{    }$\dot{}${    }''   .}    
({}\cite{buchert:averaging},                  see                 also
Appendix~\ref{sect:generalized-friedmann}):
\begin{equation} 
\label{eq:expansion-law}
3 \frac{{\ddot a}_\CD}{a_\CD} + 4\pi G\average{\varrho} -\Lambda = Q_\CD ,
\end{equation}
with  Newton's gravitational constant  $G$, the  cosmological constant
$\Lambda$, and the ``backreaction  term'' $Q_\CD$.  For $Q_\CD=0$ this
equation equals  one of the  Friedmann equations for the  scale factor
$a(t)=a_\CD(t)$ in an homogeneous  and isotropic universe with uniform
density $\varrho_{H}=\average{\varrho}$ ($Q_\CD=0$ is a necessary, but
also sufficient condition to  guarantee $a_\CD(t)=a(t)$).  As we shall
see in  the next subsection  the backreaction term $Q_\CD$  depends on
the  inhomogeneities  of  the  velocity  field inside  $\CD$,  and  is
generally not zero.

\subsection{Backreaction in Eulerian coordinates}
\label{sect:backreaction-euler}

Directly  following from averaging  Raychaudhuri's equation  we obtain
the  backreaction $Q_\CD$  depending on  the kinematical  scalars, the
expansion rate $\theta$,  the rate of shear $\sigma$,  and the rate of
vorticity $\omega$ (see Appendix~\ref{sect:generalized-friedmann}):
\begin{equation}
\label{eq:Q-kinematical-scalars}
Q_\CD = \tfrac{2}{3} \left( \average{\theta^2} - \average{\theta}^2 \right) 
+ 2 \average{\omega^2 - \sigma^2} .
\end{equation}

Often  it  is more  convenient  to work  with  the  invariants of  the
gradient  of  the  velocity  field\footnote{A  comma  denotes  partial
derivative  with respect  to  Eulerian coordinates  $\partial/\partial
x_i\equiv,i$.} as defined in Appendix~\ref{sect:invariants}:
\begin{equation}
\label{eq:backreaction-invariants}
Q_\CD = 2 \average{\inII(v_{i,j})} - \tfrac{2}{3} \average{\inI(v_{i,j})}^2 .
\end{equation}

The Eulerian approximation schemes are usually defined with respect to
a reference frame {\em comoving} with the Hubble flow.  To define such
a reference frame within an  inhomogeneous cosmology we have to assume
that for domains $\CU$  significantly larger than the averaging domain
$\CD$ the  expansion $a(t)$ is a  given solution, e.g.,  a solution of
the Friedmann equations.  Hence, using comoving coordinates we already
assume that $Q_\CU = 0$ for  very large domains $\CU$.  As we will see
from          Eq.~\eqref{eq:Q-surface-int}          (see          also
Subsect.~\ref{sect:n-body}), this can  be achieved by setting periodic
boundary conditions  on the inhomogeneous  fields at the scale  of the
domain $\CU$ (see also {}\cite{buchert:averaging} for more precision).
With  the  Hubble--parameter  $H=\dot{a}/a$  we  define  the  comoving
Eulerian  coordinates $\bq  :=\bx/a$ and  the  peculiar--velocity $\bu
:=\bv-H\bx$.
Using  the  derivative  $\partial_{q_j}u_i\equiv\partial  u_i/\partial
q_j$ with respect to comoving  coordinates we obtain for the first and
second invariants:
\begin{align}
\inI(v_{i,j}) &= 3H + \inI(u_{i,j})  = 
3H + \tfrac{1}{a}   \inI\left(\partial_{q_j}u_i\right) , \nonumber\\
\inII(v_{i,j})&= 3H^2 + 2H\inI(u_{i,j}) + \inII(u_{i,j}) \\
 & = 3H^2 + \tfrac{2H}{a} \inI\left(\partial_{q_j}u_i\right) + 
\tfrac{1}{a^2} \inII\left(\partial_{q_j}u_i\right) .
\nonumber
\end{align}
The  volume--average $\average{\CA}$ of  a tensor  field $\CA$  can be
written  as a  volume--average over  comoving domains  $\CD_q$ defined
by\footnote{The   division    $\CD/a$   is   understood   point--wise:
$\CD/a=\{\bx/a~|~\bx\in\CD\}$.} $\CD_q=\CD(t)/a(t)$  with the comoving
volume $V_q=V/a^3$:
\begin{equation}
\label{eq:average-comoving}
\average{\CA} = \frac{1}{V_q} \int_{\CD_q}\rmd^3q\ \CA .
\end{equation}
The  mass is  conserved  inside this  domain  $\CD_q$.  A  well--known
example  for such  comoving mass--conserving  domains are  the volumes
used in N--body simulations.

Now, the backreaction term in comoving coordinates reads
\begin{multline}
\label{eq:backreaction-comoving}
Q_\CD= \frac{1}{a^2} \left[ 2\ \frac{1}{V_q} \int_{\CD_q}\rmd^3q\ 
\inII\left(\partial_{q_j}u_i\right) - \right. \\
\left. -\frac{2}{3}\ \left(\frac{1}{V_q}\int_{\CD_q}\rmd^3q\ 
\inI\left(\partial_{q_j}u_i\right)\right)^2 \right],
\end{multline}
i.e.,  the form  of  $Q_\CD$ is  such  that all  Hubble  terms in  the
velocity  gradient  cancel  out:  only inhomogeneities  contribute  to
backreaction.

Using  Eqs.~\eqref{eq:v-inv-I} and  \eqref{eq:v-inv-II}  we write  the
backreaction as  a volume--average over divergences.  Hence, using the
theorem of Gauss we obtain:
\begin{multline}
\label{eq:Q-surface-int}
Q_\CD=\frac{1}{a^2} \left[ 2\ \frac{1}{V_q}\int_{\partial\CD_q} \rmd\bS \cdot
\left(\bu(\nabla_q\cdot\bu)-(\bu\cdot\nabla_q)\bu\right) - \right. \\
\left. -\frac{2}{3}\ 
\left(\frac{1}{V_q}\int_{\partial\CD_q}\rmd\bS \cdot \bu \right)^2 \right],
\end{multline}
with the  surface $\partial\CD_q$ of the comoving  domain $\CD_q$, the
surface  element  $\rmd\bS$,  and  the  comoving  differential  operator
$\nabla_q$.
From Eq.~\eqref{eq:Q-surface-int}  we directly obtain  $Q_\CD=0$ for a
domain  without  boundaries,  i.e.\  for  periodic  peculiar--velocity
fields   (see   also   the   subsection   on   N--body   simulations
{}\ref{sect:n-body}).

\subsection{Backreaction in Lagrangian coordinates}
\label{sect:lagrange-backreaction}

Let $\mbf(\cdot,t):\bX\mapsto\bx$  denote the mapping  which takes the
initial  Lagrangian   positions  $\bX$  of  fluid   elements  at  time
$\initial{t}$ to their Eulerian positions $\bx$ at time $t$; i.e. $\bx
= \mbf(\bX, t)$ and  $\bX :=\mbf(\bX,\initial{t})$.  Then the velocity
is $\bv=\dot{\mbf}(\bX,t)$,  and the  Jacobian determinant $J$  of the
mapping $\mbf$ connects the volume elements $\rmd^3x=J\rmd^3 X$.

The Lagrangian domain  $\initial{\CD}$ is connected with the  Eulerian
domain by   $\initial{\CD}=\mbf^{-1}(\CD,t)$,  as long  as   $\mbf$ is
unique.  The spatial  average of a tensor  field $\CA$ in Eulerian and
Lagrangian space are related in the following way:
\begin{equation}
\average{\CA} = \frac{1}{\laverage{J}} \frac{1}{\initial{V}}
\int_{\initial{\CD}}\rmd^3X\ J \CA 
= \frac{1}{\laverage{J}} \laverage{J\;\CA} ,
\end{equation}
with 
\begin{equation}
\label{eq:VVi-J}
\laverage{J} = \frac{V}{\initial{V}} = a_\CD^3 .
\end{equation}
The backreaction can  be expressed as a volume--average in the initial
Lagrangian   domain   $\initial{\CD}$    with   the   velocity   field
$\bv=\dot{\mbf}$ and $\bx=\mbf$:
\begin{equation}
\label{eq:Q-lagrange}
Q_\CD = 
2\ \frac{1}{\laverage{J}}\laverage{J\ \inII(v_{i,j})} 
- \frac{2}{3} \left( \frac{1}{\laverage{J}}\laverage{J\ \inI(v_{i,j})} 
\right)^2 .
\end{equation}

\subsection{A qualitative discussion}
\label{sect:qualitative-discussion}

Eqs.~\eqref{eq:expansion-law}  and  {}\eqref{eq:Q-kinematical-scalars}
show that as soon as  inhomogeneities are present, they are sources of
the  equation  governing  the  average  expansion.   Upon  integrating
Eq.~\eqref{eq:expansion-law}                 we                 obtain
({}\cite{buchert:averaging-hypothesis}    used   a    different   sign
convention for $Q_\CD$):
\begin{equation}
\label{eq:average-friedmann}
\frac{\dot{a}_\CD^2 + k_\CD}{a_\CD^2 } - \frac{8\pi G \average{\varrho}}{3}
- \frac{\Lambda}{3} = \frac{1}{3 a_\CD^2} \int_{\initial{t}}^t \rmd t'\ Q_\CD
\frac{\rmd }{\rmd s} a^2_\CD(t').
\end{equation}
$k_\CD$  enters as  an integration  constant depending  on  the domain
$\CD$.    With  $H_\CD=\dot{a}_\CD/a_\CD$   we  define   an  effective
Hubble--parameter,  and   a  dimensionless  ``kinematical  backreaction
parameter'':
\begin{equation}
\label{eq:def-omegaQ}
\Omega_Q^\CD = \frac{1}{3\ a_\CD^2 H_\CD^2} 
\int_{\initial{t}}^t\rmd t'\ Q_\CD\  2{\dot a}_\CD a_\CD
\end{equation}
in addition to the common cosmological parameters:
\begin{equation}
\label{eq:omega-local-def}
\Omega_m^\CD = \frac{8\pi G\average{\varrho}}{3H_\CD^2}, \quad 
\Omega_\Lambda^\CD = \frac{\Lambda}{3 H_\CD^2},  \quad  
\Omega_k^\CD = -\frac{k_D}{a_\CD^2 H_\CD^2} .
\end{equation}
However, here, all  $\Omega^\CD$--parameters are domain--dependent and
transformed into  fluctuating fields on the domain;  for $Q_\CD=0$ the
cosmic triangle is undistorted and the parameters acquire their global
standard     values.      Comparing     these     definitions     with
Eq.~\eqref{eq:average-friedmann} we have
\begin{equation}
\Omega_m^\CD + \Omega_\Lambda^\CD + \Omega_k^\CD + \Omega_Q^\CD = 1 .
\end{equation}
In  Friedmann--Lema\^\i{}tre  cosmologies there  is  by definition  no
backreaction: $\Omega_Q^\CD=0$.  In this case a critical universe with
$\Omega_m^\CD+\Omega_\Lambda^\CD=1$  implies $k_\CD=0$, as  usual.  If
instead  $\Omega_Q^\CD\ne0$, and  for  simplicity $\Omega_k^\CD=0$  we
have  an  additional   {\em  kinematical}  contribution  resulting  in
$1=\Omega_m^\CD+\Omega_\Lambda^\CD+\Omega_Q^\CD$.
Contrary to the standard  model, all parameters are implicit functions
of spatial scale: the  curvature parameter appearing as an integration
constant can  be different for different averaging  domains $\CD$, the
parameter  of the  cosmological constant  is  scale--dependent through
$H_\CD$.   As  we  shall  see,  $\Omega_k^\CD$  can  experience  large
changes, if initially $k_\CD$ departs (even slightly) from zero.

Let  us consider  irrotational  flows with  $\omega=0$,  which we  may
consider a good approximation until  the epoch of structure formation. 
The  averaged  shear  fluctuations  $\average{\sigma^2}\ge0$  and  the
fluctuations          in          the          expansion          rate
$\average{\theta}^2-\average{\theta^2}=
{}\average{(\theta-\average{\theta})^2}\ge0$     compete     in    the
backreaction Eq.~\eqref{eq:Q-kinematical-scalars}:
\begin{itemize}
\item
$Q_\CD$ is  positive if the  shear term dominates the  expansion term.
This leads to an accelerated structure formation at least in the early
evolutionary stages.  We then may interpret $Q_\CD$ as a ``kinematical
dark matter''.
\item
$Q_\CD$  is  negative  if  the  fluctuations  in  the  expansion  rate
dominate.  In  this case $Q_\CD$ acts like  a (positive) ``kinematical
cosmological constant'' leading to accelerated expansion.
\end{itemize}
Clearly such a ``kinematical dark matter'' will show a time dependence
different  from ordinary  (dark) matter;  also, if  $Q_\CD$  mimicks a
cosmological term,  its time--dependence  will also be  different than
that in a $\Lambda-$cosmology.

\section{Quantifying the backreaction}
\label{sect:quantifying}

In the last section we  gave different forms of the backreaction term,
and discussed qualitatively the expected consequences. Now we quantify
them.   Exact inhomogeneous  solutions  for estimating  the amount  of
backreaction are  only available for highly symmetric  models like for
the spherical  model (Subsect.~\ref{sect:spherical-symmetry}), and for
a model with  plane symmetry (Subsect.~\ref{sect:plane-symmetry}).  In
generic situations we have to rely on approximations.

\subsection{The spherical collapse}
\label{sect:spherical-symmetry}

We adopt  the usual assumptions.  The domain  of averaging $\CD=\CB_R$
is a sphere with radius $R$. The velocity $\bv$ inside $\CB_R$ is only
depending on the distance $r$ to the origin and always parallel to the
radial unit vector $\mathbf{e}_r$:
\begin{equation}
\bv=v(r)\ {\mathbf{e}}_r .
\end{equation}
Hence,  we exclude  rotational velocity  fields. Moreover  the domains
stay spherical at all times.
The  averaged first invariant  may be  obtained directly  using Gauss'
theorem:
\begin{align}
\average{\inI(v_{i,j})} & =
\frac{3}{4\pi R^3}\int_{\partial \CB_R}\rmd\bS\cdot v(r){\mathbf{e}}_r 
= 3\frac{v(R)}{R} , 
\end{align}
whereas the averages  of the second and third  invariants require some
basic       calculations.        Using       Eqs.~\eqref{eq:v-inv-II},
{}\eqref{eq:v-inv-III} and again Gauss' theorem leads to
\begin{align}
\average{\inII(v_{i,j})} = &\frac{1}{3}\average{\inI(v_{i,j})}^2 , \nonumber\\
\label{eq:spherical-I-III}
\average{\inIII(v_{i,j})} = & \frac{1}{27}\average{\inI(v_{i,j})}^3  .
\end{align}
Combining  these  terms  in  the  backreaction  $Q_{\CB_R}$  given  by
Eq.~\eqref{eq:backreaction-invariants} we get for spherical symmetry
\begin{equation}
Q_{\CB_R}^{\rm spherical} = 0,
\end{equation}
as  expected  from  the   Newtonian  analogue  of  Birkhoff's  theorem
(Newton's ``iron sphere'' theorem).

\subsection{Backreaction in the Eulerian linear approximation}

We  calculate the  backreaction in  the Eulerian  linear approximation
using Eulerian  coordinates comoving with the  background Hubble flow.
In this ansatz we assume  that there exists a well--defined background
density $\varrho_H$ and a global expansion factor $a(t)$ on very large
scales,           as           already          mentioned           in
Subsect.~\ref{sect:backreaction-euler}.    We   define   the   density
contrast $\delta=(\varrho-\varrho_H)/\varrho_H$.
Additionally we assume that the growing mode of the peculiar--velocity
field $\bu$  is dominating,  and we can  neglect rotations.   The time
evolution  of  $\bu(\bq,t)$  is   then  proportional  to  the  initial
peculiar--velocity     $\bU(\bq)=\bu(\bq,\initial{t})$    (see    also
Appendix~\ref{sect:initial}):
\begin{equation}
\label{eq:linear-peculiar-u}
\bu(\bq,t) = c(t) \bU(\bq) \;\;.
\end{equation}
With    the    irrotational    peculiar--velocity    according    to
Eq.~\eqref{eq:linear-peculiar-u} the invariants may be written as
\begin{align}
\inI\left(\partial_{q_j}u_i\right) & 
=  c(t)\ \inI\left(\partial_{q_j}U_i\right), \nonumber\\
\inII\left(\partial_{q_j}u_i\right) &
=  c^2(t)\ \inII\left(\partial_{q_j}U_i\right) .
\end{align}
A comoving domain  is approximated by\footnote{To  make this
more  rigorous we  relate $\CD_{q}=\CD(t)/a(t)$  to an  initial domain
$\initial{\CD}$ using the ``Zel'dovich approximation'' $\mbf^{Z}$ (see
Eq.~\eqref{eq:zeldovich}):
\begin{multline*}
\CD_q = \CD(t)/a(t) \approx \mbf^{Z}(\initial{\CD},t)/a(t) =\\
=\{\bq|\bq=\bX+\xi(t)\nabla_0\psi(\bX)\} 
\approx \{\bq|\bq=\bX\} = \initial{\CD},
\end{multline*}
where  the last  approximation is  valid for  $|\xi\nabla_0\psi|\ll1$. 
Padmanabhan and  Subramanian {}\cite{padmanabhan:zeldovich} discuss an
extension.}
$\CD_{q}=\CD(t)/a(t)\approx\initial{\CD}$  (the Eulerian approximation
does not strictly conserve mass).  In this approximation the volume at
time  $t$ is simply  $V=a^3\initial{V}$, and  the backreaction  in the
linear approximation  can be calculated  using Eqs.~\eqref{eq:def-ad},
{}\eqref{eq:average-comoving}, and {}\eqref{eq:backreaction-comoving}:
\begin{equation}
Q_\CD^{\rm lin} = \frac{c^2}{a^2} Q_{\initial{\CD}} 
\end{equation}
with   the  backreaction   term  evaluated   on  the   initial  domain
\begin{equation}
Q_{\initial{\CD}} = 2\ \laverage{ \inII\left(\partial_{q_j}U_i\right) } -
\frac{2}{3}\ \laverage{ \inI\left(\partial_{q_j}U_i\right) }^2 .
\end{equation}
For an Einstein--de--Sitter background $c(t)=(t/\initial{t})^{1/3}$ and
$a(t)=(t/\initial{t})^{2/3}$, and the backreaction term decays
in proportion to the global expansion:
\begin{equation}
\label{eq:Q-linear-EdS}
Q_\CD^{\rm lin} = \left(\frac{t}{\initial{t}}\right)^{-2/3} 
Q_{\initial{\CD}} .
\end{equation}

Comparing the sources  $Q_\CD^{\rm lin}$ and $\average{\varrho}^{\rm
lin}=\varrho_H(\initial{t})/a^3$ in the generalized Friedmann equation
we  find {\em  growth} of  the deviation  from the  standard Friedmann
models, on an Einstein--de--Sitter background:
\begin{equation}
\frac{Q_\CD^{\rm lin}}{4\pi G\average{\varrho}^{\rm lin}} 
\propto a^2 .
\end{equation}

With the function
\begin{equation}
h(t) = \frac{1}{a_\CD^2} 
\int_{\initial{t}}^t\rmd t'\ \frac{c^2}{a^2}\ 2{\dot a}_\CD a_\CD ,
\end{equation}
we can  write the backreaction  parameter $\Omega_Q^\CD$, as  given in
Eq.~\eqref{eq:def-omegaQ}, in the following form:
\begin{equation}
\Omega_Q^\CD =  h(t) \frac{Q_{\initial{\CD}}}{3\ H_\CD^2} ,
\end{equation}
emphasizing   the   analogy   with  a   time--dependent   cosmological
$\Lambda$--term (see Eq.~\eqref{eq:omega-local-def}).

In addition to  using the linear approximation for  the time evolution
of  the (irrotational)  velocity  field we  assume  that the  comoving
domain  $\CD_q$  is  only  weakly  deformed from  the  initial  domain
$\initial{\CD}$.  This approximation, applicable in the initial stages
of structure formation, is not needed in the Lagrangian picture, where
we have a  definite mapping from $\initial{\CD}$ to  $\CD(t)$ as given
in  Subsect.~\ref{sect:lagrange-backreaction}.  This  illustrates that
the Eulerian perturbation theory  is not well--suited to calculate the
backreaction of inherently mass conserving, i.e.\ Lagrangian, domains.

\subsection{Backreaction in the ``Zel'dovich approximation''}

Using the Lagrangian perturbation scheme one can construct approximate
solutions which are  able to trace the process  of structure formation
into the Eulerian nonlinear regime.
As  a  subclass of  the  first--order  Lagrangian perturbation  series
{}\cite{buchert:lagrangian-theory}  the  ``Zel'dovich  approximation''
{}\cite{zeldovich:fragmentation} is  frequently used and well--studied
both      theoretically      and      numerically     (see,      e.g.,
{}\cite{sahni:approximation},  and  {}\cite{buchert:lagrangian} for  a
tutorial, and refs.\ therein).

As  with   the  Eulerian   perturbation  approximation  we   assume  a
homogeneous--isotropic   background  model   on  very   large  scales,
specified by the time evolution of the global expansion factor $a(t)$.
The trajectory field in the ``Zel'dovich approximation''  is given by
\begin{equation}
\label{eq:zeldovich}
\mbf^Z(\bX,t)=a(t)\Big(\bX+\xi(t)\nabla_0\psi(\bX)\Big) .
\end{equation}
$\psi(\bX)$ is the initial displacement field, $\nabla_0$ the gradient
with   respect  to   Lagrangian   coordinates  and   $\xi$  a   global
time--dependent  function (where $\xi(a)$ is  given for  all background
models $a(t)$ in {}\cite{bildhauer:solutions}).

Given  the trajectory  field there  are two  ways of  implementing the
approximation for the effective  scale factor $a_\CD$. One is based
on  the volume  deformation of  fluid elements  due to  the trajectory
field. This results in a ``passive'' estimate of the time--dependence of
the volume measured by the Jacobian determinant:
\begin{equation}
J^Z(\bX,t) = a^3 \left(1  + \xi\initial{\inI} + \xi^2\initial{\inII} +
\xi^3\initial{\inIII} \right) ,
\end{equation}
with  the invariants of  the initial  displacement field\footnote{With
$|i$  we  denote  differentiation   with  respect  to  the  Lagrangian
coordinates $X_i$.} $\psi_{|ij}$:
\begin{equation}
\label{eq:def-invariants-displacement}
\initial{\inI}   := \inI(\psi_{|ij}),\ 
\initial{\inII}  := \inII(\psi_{|ij}),\  
\initial{\inIII} := \inIII(\psi_{|ij}) .
\end{equation}
Now $\laverage{J^Z}=:(a_\CD^{\rm kin})^3$ is given by
\begin{multline}
\label{eq:averageJZ}
(a_\CD^{\rm kin})^3 = 
a^3 \left(1 + \xi\laverage{\initial{\inI}} + 
\xi^2\laverage{\initial{\inII}}
+ \xi^3\laverage{\initial{\inIII}} \right).
\end{multline}

A  second  possibility  is  to  consider  the  dynamical  backreaction
equation  itself  and  approximate  the  backreaction  term  from  the
velocity field $\bv^Z = \dot\mbf^Z$  using $\bx=\mbf^Z$.
To ease the calculations we define the functional determinant of three
functions  $A(\bX)$,  $B(\bX)$, $C(\bX)$:
\begin{equation}
\CJ(A,B,C) := \frac{\partial(A,B,C)}{\partial(X_1,X_2,X_3)}
= \epsilon_{ijk}A_{|i} B_{|j}C_{|k} .
\end{equation}
(For  example:  the Jacobian  determinant  of  the Lagrangian  mapping
$\mbf=(f_1,f_2,f_3)$ reads  $J=\CJ(f_1,f_2,f_3)$.) Now we  express the
invariants  of  the  velocity   field  $\bv=\dot{\mbf}$  in  terms  of
functional determinants,
\begin{align}
\inI(v_{i,j})  & =
  \tfrac{1}{2 J} \epsilon_{ijk} \CJ(\dot{f}_i,f_j,f_k) ,\nonumber\\
\label{eq:jacob-invar}
\inII(v_{i,j}) & = 
  \tfrac{1}{2 J} \epsilon_{ijk} \CJ(\dot{f}_i,\dot{f}_j,f_k) ,
\end{align}
which may be verified using
\begin{equation}
v_{i,j} = \tfrac{1}{2 J}\ \epsilon_{jkl} \CJ(\dot{f}_i, f_k, f_l). 
\end{equation}

Inserting the approximation Eq.~\eqref{eq:zeldovich}  we get with
$b(t)=a(t)\xi(t)$:
\begin{align}
\CJ(\dot{f}^Z_i,f^Z_j,f^Z_k)
& = \dot{a}a^{2}\ \CJ(X_i,X_j,X_k) + 
  \dot{b}b^2\ \CJ(\psi_{|i},\psi_{|j},\psi_{|k}) \nonumber \\
& \quad + (2a\dot{a}b+a^2\dot{b})\ \CJ(X_i,X_j,\psi_{|k}) \nonumber \\
& \quad + (2a\dot{b}b+\dot{a}b^2)\ \CJ(X_i,\psi_{|j},\psi_{|k}) ,\\
\CJ(\dot{f}^Z_i,\dot{f}^Z_j,f^Z_k) 
& = \dot{a}^2a\ \CJ(X_i,X_j,X_k) + 
  \dot{b}^2b\ \CJ(\psi_{|i},\psi_{|j},\psi_{|k}) \nonumber \\
& \quad + (2\dot{a}a\dot{b}+\dot{a}^2 b)\ \CJ(X_i,X_j,\psi_{|k}) \nonumber \\
& \quad + (2\dot{a}\dot{b}b+a\dot{b}^2)\ \CJ(X_i,\psi_{|j},\psi_{|k}) .
\end{align}
Again, these functional determinants  are related to the invariants of
the      initial     displacement     field      $\psi_{|ij}$     (see
Eq.~\eqref{eq:def-invariants-displacement}):
\begin{align}
\label{eq:def-initial-inv}
\epsilon_{ijk}\CJ(X_{i},X_{j},X_{k}) & =  6  \nonumber \\
\epsilon_{ijk}\CJ(X_{i},X_{j},\psi_{|k}) & = 2 \initial{\inI},\\
\epsilon_{ijk}\CJ(X_{i},\psi_{|j},\psi_{|k}) & = 2 \initial{\inII},\nonumber \\
\epsilon_{ijk}\CJ(\psi_{|i},\psi_{|j},\psi_{|k}) &= 6 \initial{\inIII}.
\nonumber 
\end{align}
Hence, with $K:=\dot{\xi}/\xi$ and $H:=\dot{a}/a$,
\begin{align}
\label{eq:funcdet-zel-I}
\frac{1}{a^3}\ \epsilon_{ijk}\CJ(\dot{f}^Z_i,f^Z_j,f^Z_k)
= & 6 H + 2\ \initial{\inI}\ \xi(3H+K)  \nonumber \\ 
& + 2\ \initial{\inII}\ \xi^2(3H+2K) \nonumber \\
& + 6\ \initial{\inIII}\ \xi^3(H+K) , \\
\label{eq:funcdet-zel-II}
\frac{1}{a^3}\ \epsilon_{ijk}\CJ(\dot{f}^Z_i,\dot{f}^Z_j,f^Z_k)
= & 6 H^2 + 2\ \initial{\inI}\ \xi (3H^2+2HK) \nonumber \\ 
& + 2\ \initial{\inII}\ \xi^2 (3H^2+4HK+K^2) \nonumber \\
& + 6\ \initial{\inIII}\ \xi^3 (H^2+2HK+K^2) .
\end{align}
Using     Eqs.~(\ref{eq:funcdet-zel-I},     {}\ref{eq:funcdet-zel-II})
inserted into Eq.~\eqref{eq:jacob-invar} we can calculate the invariants
of $v_{i,j}$ as they appear in Eq.~\eqref{eq:Q-lagrange}.
Combined with Eq.~\eqref{eq:averageJZ} the backreaction term separates
into its time--evolution given  by $\xi(t)$ and the spatial dependence
on  the  initial  displacement   field  given  by  averages  over  the
invariants $\initial{\inI}, \initial{\inII}, \initial{\inIII}$:
\begin{multline}
\label{eq:Q-full-zel}
Q^Z_\CD = \frac{\dot{\xi}^2}{
\left(1 + \xi\laverage{\initial{\inI}} + \xi^2\laverage{\initial{\inII}}
 + \xi^3\laverage{\initial{\inIII}} \right)^{2}} \times \\[1ex]
\Big[ 
\left( 2\laverage{\initial{\inII}} - 
  \tfrac{2}{3}\laverage{\initial{\inI}}^2 \right) +
\xi \left(6\laverage{\initial{\inIII}}-
  \tfrac{2}{3}\laverage{\initial{\inI}}\laverage{\initial{\inII}}\right)\\
+ \xi^2 \left(2\laverage{\initial{\inI}}\laverage{\initial{\inIII}}
  -\tfrac{2}{3}\laverage{\initial{\inII}}^2 \right)  \Big] .
\end{multline}
The  numerator of  the first  term is  global and  corresponds  to the
linear   damping   factor;   in   an   Einstein--de--Sitter   universe
$\dot{\xi}^2 \propto a^{-1}$.  The denominator  of the first term is a
volume  effect,  whereas the  second  term  in  brackets features  the
initial backreaction as a leading term.

In the  following we will stick to  an Einstein--de--Sitter background
with                  $a(t)=(t/\initial{t})^{2/3}$                 and
$\nabla_0\psi(\bX)=3/2\bU(\bX)\initial{t}$
{}\cite{zeldovich:fragmentation,buchert:lagrangian-theory}.    $\xi(t)$
equals     $a(t)-1$,     since    $\mbf(\bX,\initial{t})=\bX$     with
$a(\initial{t})=1$.

In the  early stages of  structure formation with $\xi(t)\ll1$  we get
\begin{equation}
Q^Z_\CD \approx 
\left(\frac{t}{\initial{t}}\right)^{-\frac{2}{3}}\ Q_\initial{\CD},
\end{equation}
identical to the early  evolution in the Eulerian linear approximation
Eq.~\eqref{eq:Q-linear-EdS}.   Moreover, for  $t=\initial{t}$  we have
$Q^Z_\CD=Q_\initial{\CD}$                consistent               with
Eq.~\eqref{eq:expansion-law}.
For  late times  and  positive invariants  $\laverage{\initial{\inI}},
\laverage{\initial{\inII}},\laverage{\initial{\inIII}}>0$           the
dominating  contribution is  proportional  to $\dot{\xi}^2/\xi^4$  and
$Q^Z_\CD$  decays.  However,  for negative  invariants,  the $Q^Z_\CD$
term diverges  at some time  when $1 +  \xi\laverage{\initial{\inI}} +
\xi^2\laverage{\initial{\inII}}   +  \xi^3\laverage{\initial{\inIII}}$
passes through  zero.  This is  the signature of  pancake--collapse on
the  scale of  the  domain.  A  variety  of complex  phenomena may  be
expected for averaged invariants with different signs.

\subsection{An exact reference solution: the plane collapse}
\label{sect:plane-symmetry}

The    ``Zel'dovich     approximation''    is    an     {\em    exact}
three--di\-men\-sio\-nal  solution   to  the  Newtonian   dynamics  of
self--gra\-vi\-ta\-ting  dust--matter   for  initial  conditions  with
$\inII(\psi_{|ij})=0=\inIII(\psi_{|ij})$       at      each      point
{}\cite{buchert:class}.   This ``locally  one--dimensional''  class of
motions contains as a  sub case the globally plane--symmetric solution
{}\cite{zentsova:evolutionII}.  Eq.~\eqref{eq:Q-full-zel} reduces to
\begin{equation}
\label{eq:Q-plane-collapse}
Q^{\rm plane}_\CD  = -\frac{2}{3}\
\frac{\dot{\xi}^2 \laverage{\initial{\inI}}^2}{
(1 + \xi\laverage{\initial{\inI}})^2} .
\end{equation}
We  compare the  two  exact  solutions, the  plane  and the  spherical
model in Subsect.~\ref{sect:comp-plane-spherical}.
For negative $\laverage{\initial{\inI}}$, corresponding to over--dense
regions (Eq.~\eqref{eq:Delta-I}), $Q^{\rm  plane}_\CD$ is diverging at
some time when $1  + \xi(t)\laverage{\initial{\inI}}$ approaches zero. 
Our special  initial conditions  imply a one--dimensional  symmetry of
inhomogeneities, and the diverging  $Q^{\rm plane}_\CD$ is supposed to
mimick    the   highly   anisotropic    pancake   collapse    in   the
three--dimensional situation.

\subsection{Zel'dovich's approximation and the spherical collapse}
\label{sect:zeldovich-spherical-exact}

In      the      situation      of      a      spherical      collapse
(Subsect.~\ref{sect:spherical-symmetry})   we  have   seen   that  the
averaged invariants obey relations resulting in
\begin{equation}
\baverage{\initial{\inII}} = \frac{1}{3} \baverage{\initial{\inI}}^2 ,
\quad \text{and} \quad
\baverage{\initial{\inIII}} = \frac{1}{27} \baverage{\initial{\inI}}^3 .
\end{equation}
Inserting this into the backreaction term Eq.~\eqref{eq:Q-full-zel} we
obtain   $Q^Z_{\CB_R}=0=Q^{\rm    spherical}_{\CB_R}$.    Hence,   the
``Zel'dovich approximation''  together with the  generalized Friedmann
equation results  also in an exact  solution for the  evolution of the
effective scale  factor $a_\CD$, if restricted to  the special initial
conditions $\bU=U(r)\ \mathbf{e}_r$.

\section{The evolution of cosmological parameters}
\label{sect:evolution-cosmo}

The backreaction  term itself decays  in the linear  approximation and
also at early stages  in the ``Zel'dovich approximation''. However, to
quantify the deviations  from the behavior of the  scale factor in the
standard  model,  the  different   strength  of  the  sources  in  the
generalized Friedmann  equation has to be compared.   As already noted
in the  Eulerian linear approximation,  the dimensionless contribution
to backreaction (as compared to the global mean density) grows.

\subsection{Evolution of the scale factor $a_\CD$}

We proceed as follows. We use the backreaction terms calculated in the
preceding  sections, insert  them into  the generalized  expansion law
Eq.~\eqref{eq:expansion-law}, and solve  for $a_\CD(t)$.  We determine
generic    initial   conditions    numerically   as    summarized   in
Appendix~\ref{sect:initial}.

Assuming  an Einstein--de--Sitter  background we  subtract  a standard
Friedmann  equation  $3\tfrac{{\ddot  a}}{a}+4\pi G\varrho_H=0$,  from
Eq.~\eqref{eq:expansion-law}, use $\varrho_H=\tfrac{3H^2}{8\pi G}$ and
obtain the following differential equation for $a_\CD(t)$:
\begin{equation}
\label{eq:evolution-ad}
3\left( \frac{{\ddot a_\CD}}{a_\CD}-\frac{{\ddot a}}{a}\right) +
\frac{3}{2} \left(\frac{{\dot a}}{a}\right)^2 \average{\delta} 
= Q_\CD \;\;,
\end{equation}
with         $\average{\delta}=(\average{\varrho}-\varrho_H)/\varrho_H$
specifying the averaged density  contrast in $\CD$.  For $Q_\CD=0$ and
$\average{\delta}=0$  this  equation   simply  states  that  the  time
evolution of  a domain follows the global  expansion, $a_\CD(t)=a(t)$. 
The  averaged  density  contrast  is  related to  the  averaged  first
invariant      $\laverage{\initial{\inI}}$       as      given      in
Eq.~\eqref{eq:Delta-I}.            For          $Q_\CD=0$          and
$1+\average{\delta}=\frac{\average{\varrho}  a^3}{\varrho_H  a_\CD^3}$
the evolution of  $a_\CD$ is still of Friedmann type,  but with a mass
different from the homogeneous background mass.  An important sub case
with $Q_\CD=0$ and $\average{\delta}\ne0$ is the well--known spherical
top--hat  model {}\cite{peebles:lss}.   In Eq.~\eqref{eq:evolution-ad}
there are  two sources determining  the deviations from  the Friedmann
acceleration, the over/under--density {\em and} the backreaction term:
the   evolution   of   a   spatial   domain   is   triggered   by   an
over/under--density     {\em    and}     velocity     fluctuations.    
Eq.~\eqref{eq:evolution-ad}   can   be   interpreted  as   a   natural
generalization of the top--hat model.
Engineer  et al.\  {}\cite{engineer:nonlinear} also  incorporate shear
and  angular momentum to  formulate and  study generalizations  of the
spherical top--hat model.  Their emphasis, however, is focussed on the
improvement of the virialization and other clustering arguments.

To solve this differential  equation we specify the initial conditions
according     to          the  background     model:      
\begin{equation}
a(t) = \left(\frac{t}{\initial{t}}\right)^{2/3} \text{ with } 
\initial{t} 
= \frac{2 }{3 H_0 (1+\initial{z})^{3/2}}. 
\end{equation}
$H_0$ is the  present value of the Hubble--parameter of the background
model, and $\initial{z}$ the redshift where we start our calculations.
With $H_0=50{\rm km\ s}^{-1} {\rm Mpc}^{-1}$ and $\initial{z}=200$ the
initial starting time is $\initial{t}=4.6\ 10^{-3}{\rm Gy}$.
The  scale  factor  $a_\CD$ of  the  domain  is  chosen to  match  the
background scale factor $a$ at time $\initial{t}$:
\begin{equation}
a_\CD(\initial{t}) = a(\initial{t}) = 1 .
\end{equation}
We  stop  at  the present  epoch   $z_0=0$ and $t_0=13{\rm  Gy}$, with
$a(t_0)=201$.
Following  directly from  Eq.~\eqref{eq:theta-HD},  ${\dot a_\CD}$  is
given by
\begin{equation}
\label{eq:consistent}
{\dot a_\CD}(\initial{t}) 
= {\dot a}(\initial{t}) \left(1+\tfrac{1}{3}\laverage{\initial{\inI}}\right)
= \frac{2}{3\initial{t}} \left(1+\tfrac{1}{3}\laverage{\initial{\inI}}\right).
\end{equation}
Often  ${\dot a_\CD}(\initial{t})={\dot  a}(\initial{t})$  is used  to
specify the initial conditions for the investigation of collapse times
within  the spherical model.   Numerical tests  showed that  for these
inconsistent  initial conditions  the collapse  is delayed  as already
discussed by Bartelmann et al.~\cite{bartelmann:timescales}.

For the  numerical integration we used the  Runge--Kutta routine given
in Press et al.~\cite{press:recipes}.  Identical results were obtained
with the {\tt NDSolve} routine of {}\textsc{Mathematica}.

\subsubsection{Initial conditions}

In  Sect.~\ref{sect:quantifying} we expressed  the term  $Q_\CD$ using
globally  given  time--functions   $a(t)$,  $\xi(t)$,  and  using  the
volume--averaged        invariants        $\laverage{\initial{\inI}}$,
$\laverage{\initial{\inII}}$,  $\laverage{\initial{\inIII}}$  for  the
initial    domain   $\initial{\CD}$.     The    relation   of    these
volume--averaged invariants to the initial power spectrum is discussed
in   Appendix~\ref{sect:initial}   in    detail   using   the   linear
approximation.   We  determine  the  initial values  of  the  averaged
invariants  for a  spherical  initial domain  $\CB_R$  of radius  $R$,
assuming a  Gaussian initial density field,  with a Cold--Dark--Matter
(CDM) power spectrum.

The  ensemble  mean  $\BE$,   i.e.\  the  expectation  value,  of  the
volume--averaged  invariants vanishes  in conformity  with  our global
set--up:
\begin{equation}
\label{eq:I=0=II=0=III}
\BE\left[\baverage{\initial{\inI}}\right] = 
\BE\left[\baverage{\initial{\inII}}\right] = 
\BE\left[\baverage{\initial{\inIII}}\right] = 0.
\end{equation}
However, for a specific domain, any of the volume--averaged invariants
may  be positive  or negative.   These invariants  fluctuate  with the
variance,                                                         e.g.,
$\sigma_{\inI}^2(R)=\BE\left[\baverage{\initial{\inI}}^2\right]$.    We
consider           one--$\sigma$          fluctuations,          e.g.\ 
$\baverage{\initial{\inI}}=\pm\sigma_{\inI}(R)$,  and  also two--  and
three--$\sigma$  fluctuations  of   the  averaged  invariants  in  our
calculations of the time evolution of $a_\CD(t)$.
Clearly, these fluctuations depend on  the shape of the power spectrum
and the radius of the initial domain.  In Table.~\ref{table:sigmas} we
give  the  values  for   spherical  domains  of  different  sizes,  as
calculated in Appendix~\ref{sect:initial} for  the standard CDM model. 
Moreover,  we  could show  that  the  volume--averaged invariants  are
uncorrelated  for  a  Gaussian  initial  density  field,  and  may  be
specified independently.
\begin{table}
\caption{    The    mean    mass   fluctuations    $\sigma_{\inI}(R)$,
$\sigma_{\inII}(R)$, and $\sigma_{\inIII}(R)$ for an initial domain of
radius $R$  are given,  calculated using the  CDM power  spectrum.  To
make  these numbers  more accessible,  also the  linearly extrapolated
mean  fluctuations $a(t_0)\sigma_{\inI}(R)$ for  a domain  with scaled
radius    $a(t_0)R$    at   present    time    are   given    (compare
Fig.~\ref{fig:domains}).}
\label{table:sigmas}
\begin{tabular}{lrrrrrrr}
$aR$\ [Mpc] 
& 5 & 16 & 50  & 100 & 251  & 503   & 1005 \\[1ex]
\hline\\
$a\sigma_{\inI}$   
& 2.6 & 1.0 & 0.27  & 0.1 &  0.025 & 0.0068   & 0.0019 \\
$R$\ [Mpc] 
& 0.025 & 0.08 & 0.25  & 0.5 & 1.25 & 2.5   & 5 \\
$\sigma_{\inI}$\ $\times 10^{3}$
& 13 & 5.0 & 1.4  & 0.51 &  0.13 & 0.03   & 0.01 \\
$\sigma_{\inII}$\ $\times 10^{6}$
& 90 & 20 & 2  & 0.4 &  0.07 & 0.02   & 0.004 \\
$\sigma_{\inIII}$\ $\times 10^{12}$
& 81000 & 4600 & 100  & 5 & 0.08  & 0.001   &  0.00004\\
\end{tabular}
\end{table}

\subsubsection{The plane and spherical collapse}
\label{sect:comp-plane-spherical}

We calculate the evolution of the scale factor $a_\CD(t)$ according to
Eq.~\eqref{eq:evolution-ad}  using  the  backreaction term  calculated
within  the ``Zel'dovich approximation''  {}\eqref{eq:Q-full-zel} with
$a(t)=(t/\initial{t})^{2/3}$ and $\xi(t)=a(t)-1$.

\begin{figure}
\begin{center}
\epsfig{figure=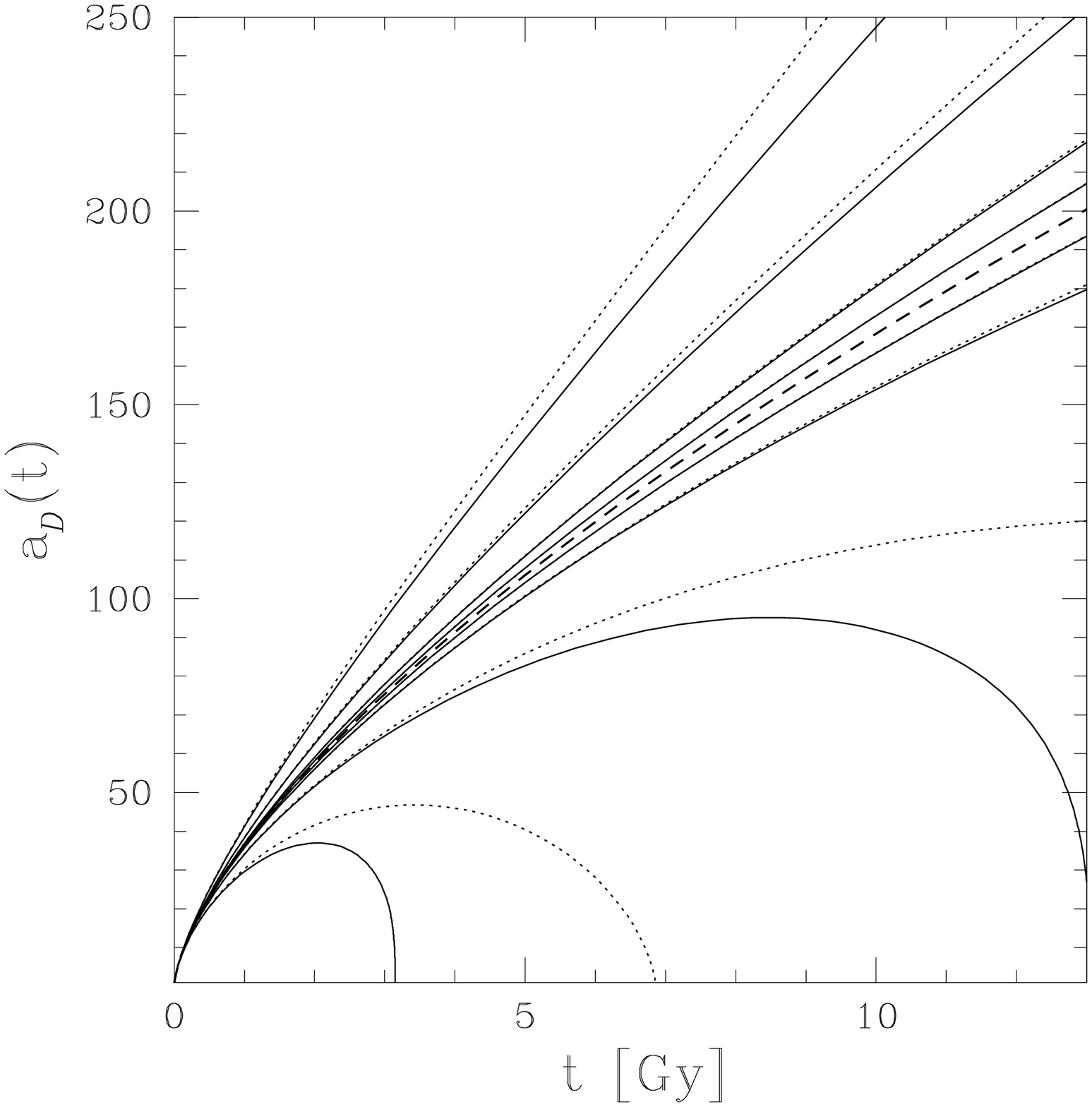,width=8cm}
\end{center}
\caption{\label{fig:plane-spherical}    The     curves show  the
  scale factor  $a_\CD(t)$ in the spherical collapse  (dotted) and the
  plane               collapse               (solid)               for
  $\baverage{\initial{\inI}}\in\{-13,-5.0,-1.4,-0.51,
  0.51,1.4,5.0,13\}\times10^{-3}$,  bending   up  successively.  These
  values correspond to one--$\sigma$ fluctuations of spherical domains
  with   a  scaled   radius  of   $a(t_0)R\in\{5,16,50,100\}$Mpc  (see
  Table~\ref{table:sigmas}).   The homogeneous background  solution is
  given as the dashed line.}
\end{figure}

First       we       consider       initial      conditions       with
$\baverage{\initial{\inII}}=0=\baverage{\initial{\inIII}}$          and
$\baverage{\initial{\inI}}\ne0$,    the   plane    collapse   solution
Eq.~\eqref{eq:Q-plane-collapse}.                We              choose
$\baverage{\initial{\inI}}\ne0$  comparable to  the  mean fluctuations
$\sigma_{\inI}(R)$   in  initially   spherical  domains   with  radius
$R\in\{0.025,0.08,0.25,0.5\}$Mpc  (a  scaled,  present--day radius  of
$a(t_0)R\in\{5,16,50,100\}$Mpc) as given in Table~\ref{table:sigmas}.
From Fig.~\ref{fig:plane-spherical} it can  be seen that the evolution
of  the scale  factor  $a_\CD(t)$ may  significantly  differ from  the
background expansion  $a(t)$ for an initial  domain radius $R=0.25$Mpc
(a scaled  radius of $a(t_0)R=50$Mpc).  For small  domains with larger
$\baverage{\initial{\inI}}$ the effect is more pronounced.
For negative  $\baverage{\initial{\inI}}$ (an over--dense  region) the
region shows  a pancake  collapse with $Q_\CD\rightarrow\infty$,  as a
result of the one--dimensional symmetry in the initial conditions.
For   initial  domains   with   $R\ge2.5$Mpc  (a   scaled  radius   of
$a(t_0)R\ge503$Mpc)   the  value  of   $\baverage{\initial{\inI}}$  is
typically so  small that the $a_\CD(t)$ closely  follows the Friedmann
solution until present.
In the following we compare the two inhomogeneous exact solutions, the
plane  and  spherical collapse,  disentangling  the  influence of  the
backreaction term $Q_\CD$ from the influence of the over--density.

As  shown in Subsect.~\ref{sect:spherical-symmetry},  the backreaction
$Q_\CD$  is  zero for  a  spherical  domain  with spherical  symmetric
velocity  field.   However,  this   domain  may  still  be  over--  or
under--dense compared to the background density.  The evolution of the
scale factor    of    this    domain    is   then    determined    by
Eq.~\eqref{eq:evolution-ad} with $Q_\CD=0$ and $\average{\delta}\ne0$,
which is exactly the spherical collapse model {}\cite{peebles:lss}.

In   Fig.~\ref{fig:plane-spherical}  we   compare  the   evolution  of
$a_\CD(t)$ in  this spherical collapse  model with the  plane collapse
model.   For  $|\baverage{\initial{\inI}}|=1.4$,  typical for  domains
with a scaled  radius of $a(t_0)R=50$Mpc, both models  give nearly the
same evolution.  However for increasingly over-dense regions (negative
$\baverage{\initial{\inI}}$) the  plane model decouples  and collapses
significantly earlier than the  spherical model. Also the expansion of
under--dense regions  is slowing down  in the plane compared  with the
spherical collapse model. These  results are in agreement with earlier
results  on collapse--times  of structure  in  Lagrangian perturbation
schemes         {}\cite{alimi:collisionless,moutarde:precollapse,%
bartelmann:timescales,buchert:performance,karakatsanis:temporal}.

\subsubsection{Generic initial data}

\begin{figure}
\begin{center}
\epsfig{figure=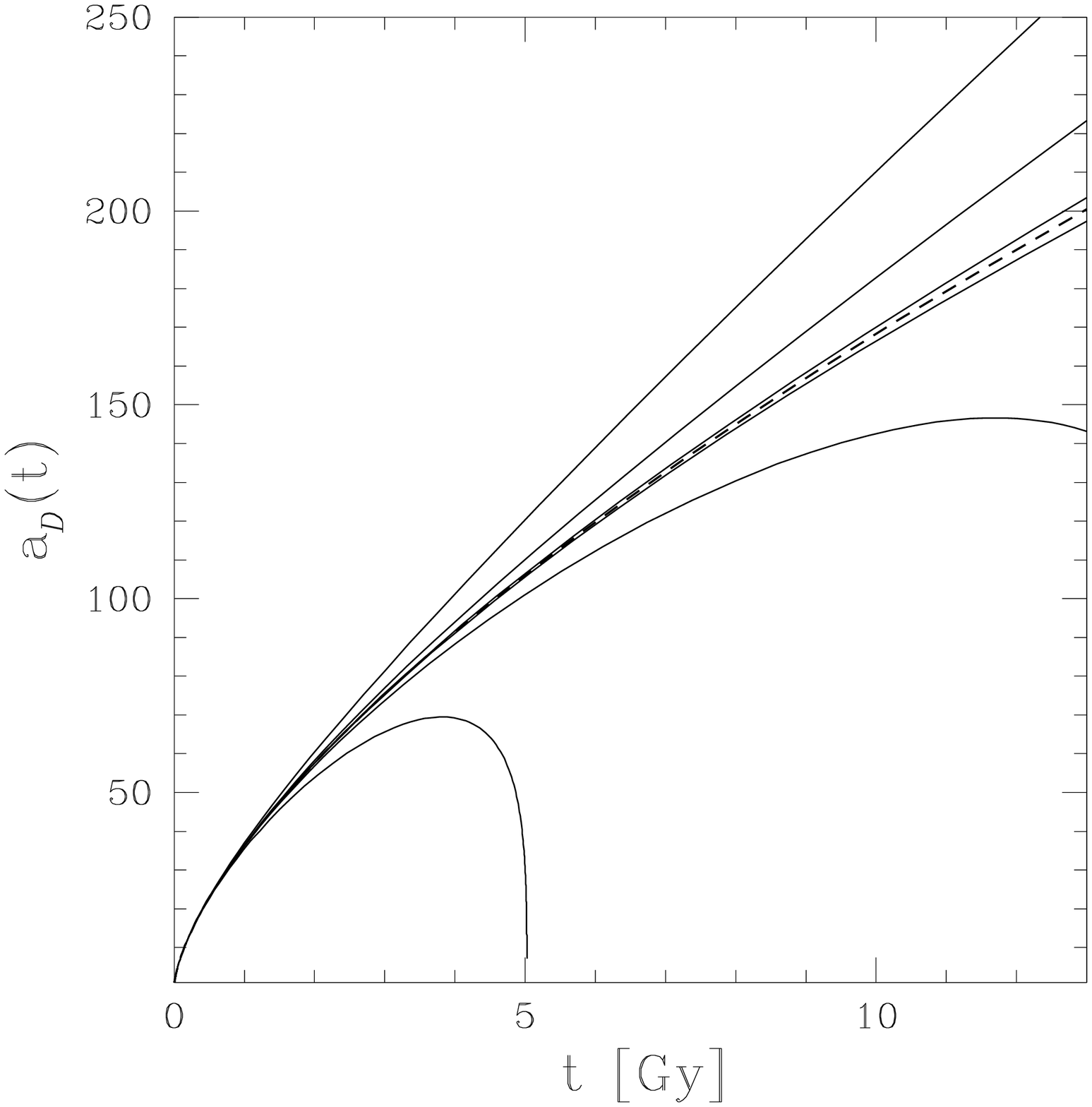,width=8cm}
\end{center}
\caption{\label{fig:aD-zel-I0II}  The curves shown are the evolution  
  of  the   scale  factor  $a_\CD(t)$  for   initial  conditions  with
  $\baverage{\initial{\inI}}=0=\baverage{\initial{\inIII}}$         and
  $\baverage{\initial{\inII}}\in\{-90,-20,-2,2,20,90\}\times10^{-6}$
  shown  as the  solid lines  bending up  successively.   These values
  correspond to one--$\sigma$ fluctuations of spherical domains with a
  scaled radius of \{5,16,50\}Mpc (see Table~\ref{table:sigmas}).  The
  homogeneous background solution is given as the dashed line.}
\end{figure}

The behavior of $a_\CD(t)$ becomes more interesting if we consider the
other  two   invariants,  going  beyond  the   exact  solutions.   For
simplicity we choose  $\baverage{\initial{\inI}}=0$ corresponding to a
domain with  $\baverage{\delta}=0$ matching the  background density at
initial time $\initial{t}$.  The  effect of a nonzero second invariant
$\baverage{\initial{\inII}}\ne0$         is         visible         in
Fig.~\ref{fig:aD-zel-I0II}.   Already  in  a  domain with  an  initial
radius of  $R=0.25$Mpc (a scaled radius  of $a(t_0)R\approx50$Mpc) the
$a_\CD$ may deviate from the Friedmann solution.  Although we consider
volumes  with no  initial  over--density and  with only  one--$\sigma$
fluctuations  of $\baverage{\initial{\inII}}$,  domains  of a  current
size of $a(t_0)R\approx16$Mpc may collapse.

\begin{figure}
\begin{center}
\epsfig{figure=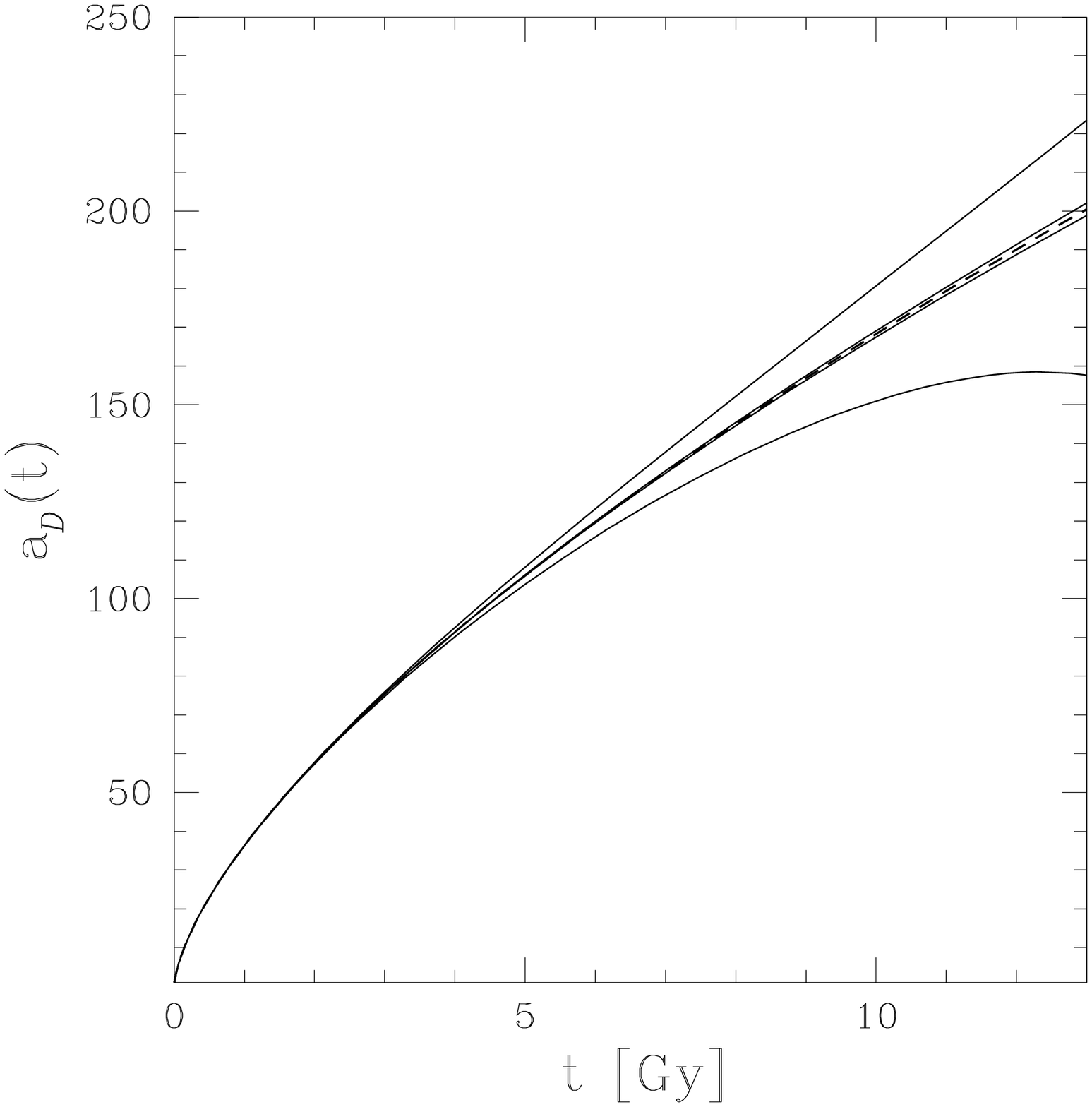,width=8cm}
\end{center}
\caption{\label{fig:aD-zel-I0II0III} The evolution of the scale factor
  $a_\CD(t)$         for        initial         conditions        with
  $\baverage{\initial{\inI}}=0=\baverage{\initial{\inII}}$          and
  $\baverage{\initial{\inIII}}\in\{-81,-4.6,4.6,81\}\times10^{-15}$
  shown  as the  solid lines  bending up  successively.   These values
  correspond to one--$\sigma$ fluctuations of spherical domains with a
  present    day    scaled    radius    of   5    and    16Mpc    (see
  Table~\ref{table:sigmas}).   The homogeneous background  solution is
  given as the dashed line.}
\end{figure}

The influence  of the  third invariant on  the evolution of  the scale
factor is  small (in  the temporal range  considered) compared  to the
influence          of          $\baverage{\initial{\inI}}$         and
$\baverage{\initial{\inII}}$.  To illustrate  this we consider domains
with    $\baverage{\initial{\inI}}=0=\baverage{\initial{\inII}}$   and
one--$\sigma$  fluctuations of $\baverage{\initial{\inIII}}$.   For an
initial  domain of  radius  $0.025$Mpc  (a scaled  radius  of 5Mpc)  a
significant  influence  is  visible (Fig.~\ref{fig:aD-zel-I0II0III}).  
Already for an initial domain  of radius $0.08$Mpc (a scaled radius of
16Mpc) the influence is small, becoming negligible on larger scales.

\begin{figure}
\begin{center}
\epsfig{figure=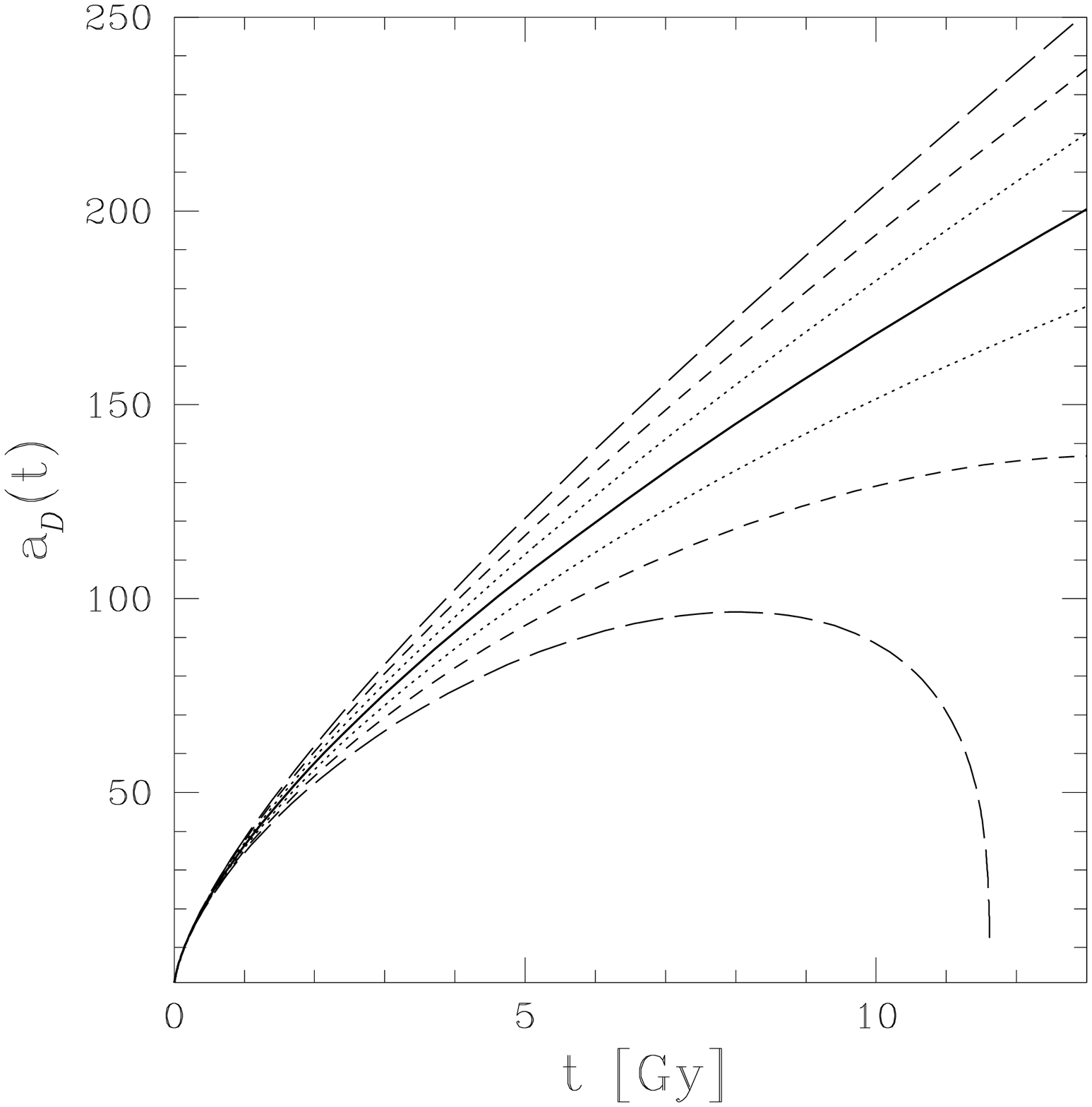,width=8cm}
\end{center}
\caption{\label{fig:aD-zel-50Mpc}  The evolution  of the  scale factor
  $a_\CD(t)$ for spherical domains with  a scaled radius of 50Mpc.  We
  considered one--$\sigma$ (dotted  line), two--$\sigma$ (short dashed
  line),  and  three--$\sigma$  (long  dashed  line)  fluctuations  of
  $\baverage{\initial{\inI}}$,    $\baverage{\initial{\inII}}$,    and
  $\baverage{\initial{\inIII}}$  (see  Table~\ref{table:sigmas}).  The
  evolution   shown  are   obtained   for  $\baverage{\initial{\inI}},
  \baverage{\initial{\inII}},            \baverage{\initial{\inIII}}<0$
  (over--dense)             and            $\baverage{\initial{\inI}},
  \baverage{\initial{\inII}},            \baverage{\initial{\inIII}}>0$
  (under--dense).  The homogeneous background solution is given by the
  solid line.}
\end{figure}

Up to now we only considered domains with initial conditions chosen as
one--$\sigma$  fluctuations   in  either  $\baverage{\initial{\inI}}$,
$\baverage{\initial{\inII}}$,  or  $\baverage{\initial{\inIII}}$.   As
can  be seen  from  Fig.~\ref{fig:aD-zel-50Mpc} the  evolution of  the
scale factor is strongly depending on the initial conditions, even for
a domain with initial radius 0.25Mpc (a scaled radius of 50Mpc), if we
also consider  two--$\sigma$ or three--$\sigma$  fluctuations.  Such a
domain may even collapse.

\begin{figure}
\begin{center}
\epsfxsize=8cm
\epsfig{figure=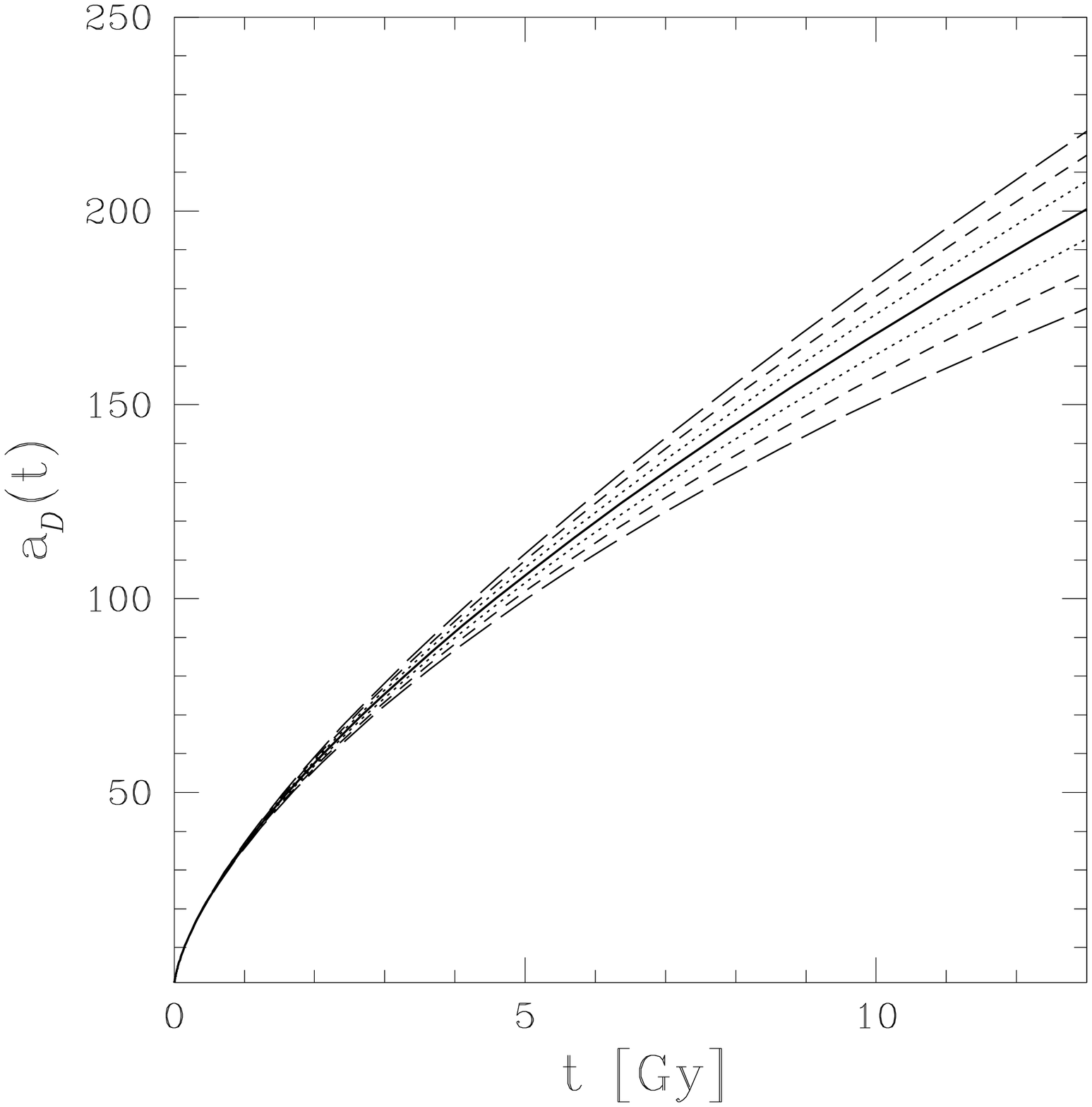,width=8cm}
\epsfig{figure=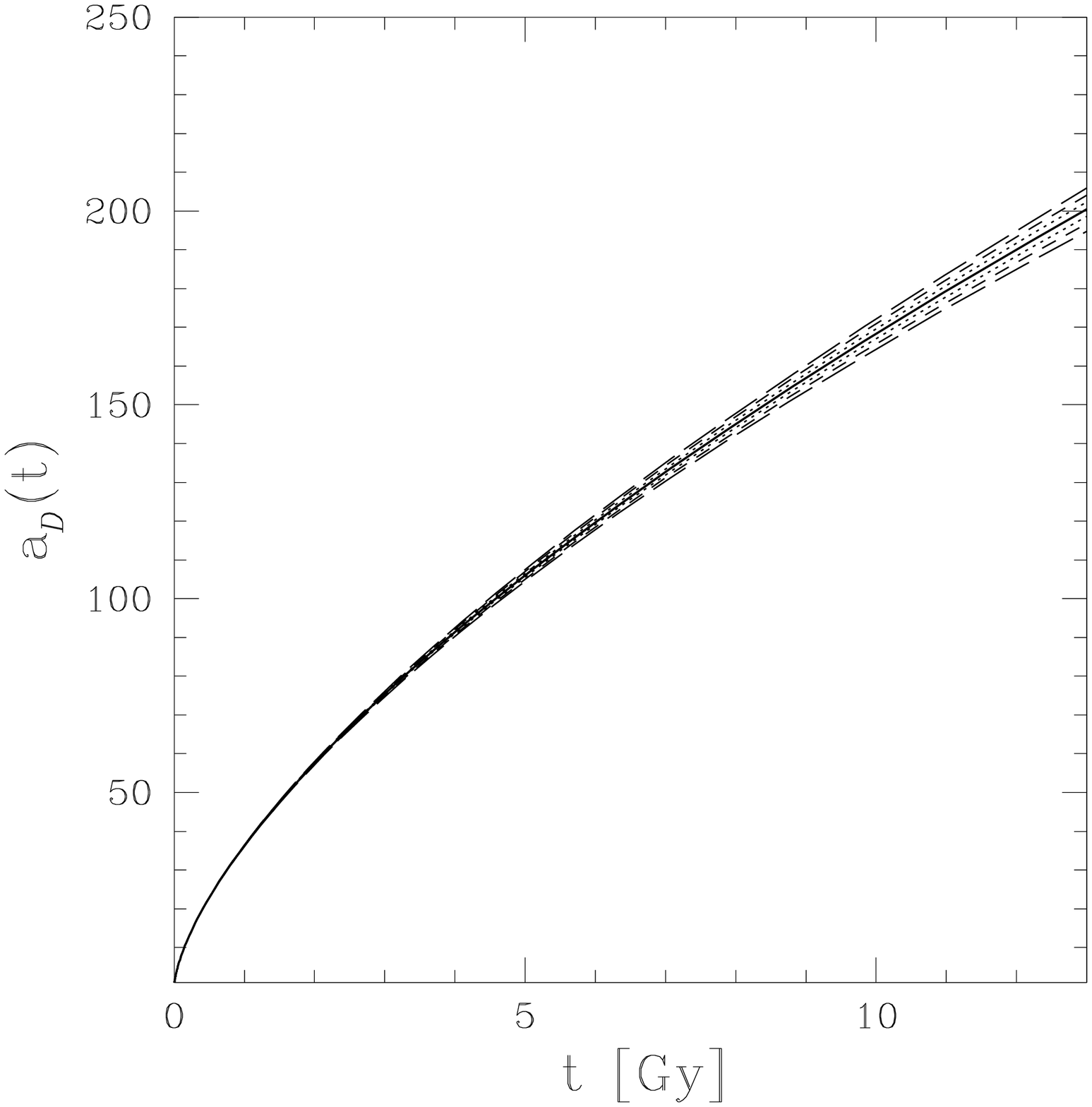,width=8cm}
\end{center}
\caption{\label{fig:aD-zel-100-250Mpc}  The  evolution  of  the  scale
  factor  $a_\CD(t)$ for  spherical domains  with a  scaled  radius of
  100Mpc (upper plot) and 250Mpc (lower plot). The same conventions as
  in Fig.~\ref{fig:aD-zel-50Mpc} are used.}
\end{figure}

Using  the present approach  we already  assume that  the backreaction
term gives a negligible contribution on the largest scale.  Hence, for
increasing  size of  the initial  domain  the evolution  of the  scale
factor goes  more and more  conform with the Friedmann  evolution (see
Fig.~\ref{fig:aD-zel-100-250Mpc}).  Still  for a domain  with a scaled
radius of 100Mpc  deviations of the order of 15\%  in the scale factor
at  present time  may  be  obtained for  initial  fluctuations at  the
three-$\sigma$ level.  The scaled  volume of such a domain corresponds
to a cube  with $\approx160$Mpc side length --  still a typical volume
used in  $N$--body simulations.  For  domains with a scaled  radius of
250Mpc ($\approx400$Mpc side  length of a cube) we  may get deviations
of the  order of  3\% from the  Friedmann value $a(t_0)$,  for initial
fluctuations      at      the      three-$\sigma$      level.       In
Sects.~\ref{sect:evolution-HD-qD}                                   and
{}\ref{sect:evolution-density-par} we  will see that  the influence on
the domains'  Hubble-- and deceleration--parameter  as well as  on its
density parameters is more pronounced, because these parameters invoke
the time--derivatives of $a_\CD(t)$.

\subsubsection{Comparison with the Eulerian linear approximation}

\begin{figure}
\begin{center}
\epsfig{figure=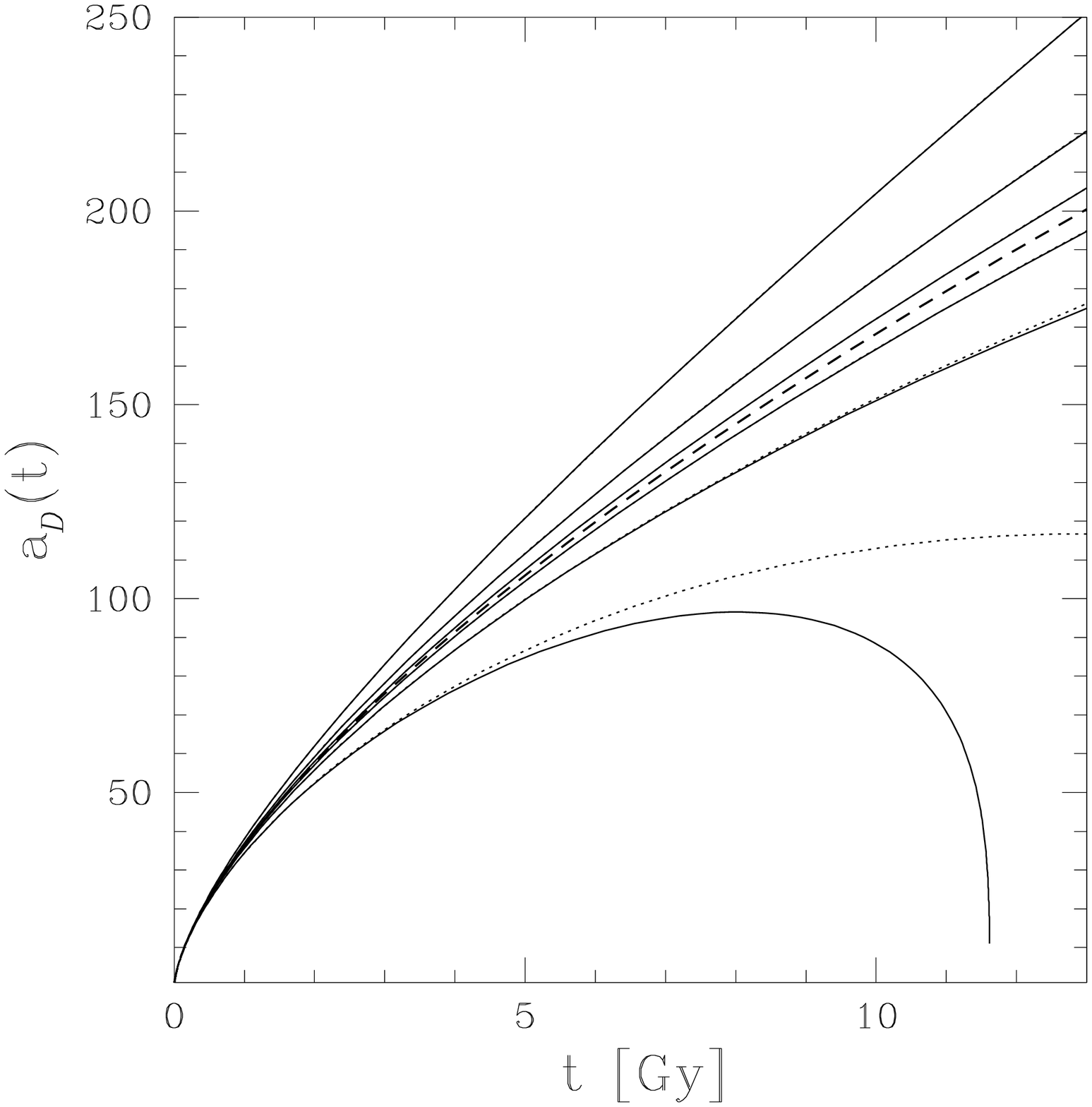,width=8cm}
\end{center}
\caption{\label{fig:aD-zel-lin}  The  evolution  of the  scale  factor
  $a_\CD(t)$ for three--$\sigma$ initial conditions as already used in
  Fig.~\ref{fig:aD-zel-100-250Mpc}.   The curves  are  for over--dense
  domains with  a scaled radius  of 50, 100, 250Mpc,  and under--dense
  domains  with  a  scaled  radius  of 250,  100,  50Mpc,  bending  up
  successively.   The evolution  of $a_\CD^Z$  is given  by  the solid
  lines, the evolution of $a_\CD^{\rm  lin}$ by the dotted lines.  The
  evolution according to the Friedmann solution is given by the dashed
  line.}
\end{figure}

Using the backreaction term~\eqref{eq:Q-linear-EdS} in the generalized
evolution  equation~\eqref{eq:evolution-ad}   we  calculate  the  time
evolution of $a_\CD^{\rm lin}(t)$ in the Eulerian linear approximation.
In  Fig.~\ref{fig:aD-zel-lin} we  compare the  evolution of  the scale
factor in  the ``Zel'dovich approximation'' with the  evolution in the
Eulerian    linear   approximation.     Even   though    we   consider
three--$\sigma$  initial  conditions,  the evolution  of  under--dense
domains with  a scaled  radius larger than  50Mpc is  described nearly
perfectly  by  the  Eulerian  linear approximation.   For  over--dense
domains  the  concordance  with  the ``Zel'dovich  approximation''  is
reasonable for domains  with a scaled radius larger  than 100Mpc.  The
Eulerian linear approximation fails to describe a collapsing domain.
However, we see  that the backreaction term is  not only important for
small domains,  where the nonlinear  evolution plays a  dominant role,
but also for large domains within the realm of linear theory.
We  emphasize  that  these  remarks  are concerned  with  the  average
performance; linear  theory does not  predict correctly where  mass is
moved,  as  cross--correlation   tests  with  Lagrangian  perturbative
schemes  and N--body  runs (even  if smoothed  on larger  scales) have
shown {}\cite{melott:testing}.

\subsubsection{Evolution of the Hubble-- and deceleration--parameter}
\label{sect:evolution-HD-qD}

Not only the evolution of  the scale factor itself is interesting, but
also  the derived quantities  like the  Hubble--parameter $H_\CD={\dot
  a}_\CD/a_\CD$  and the  deceleration  parameter $q_\CD=-\frac{{\ddot
    a}_\CD/a_\CD}{H_\CD^2}$ are of specific interest.

\begin{figure}
\begin{center}
\epsfig{figure=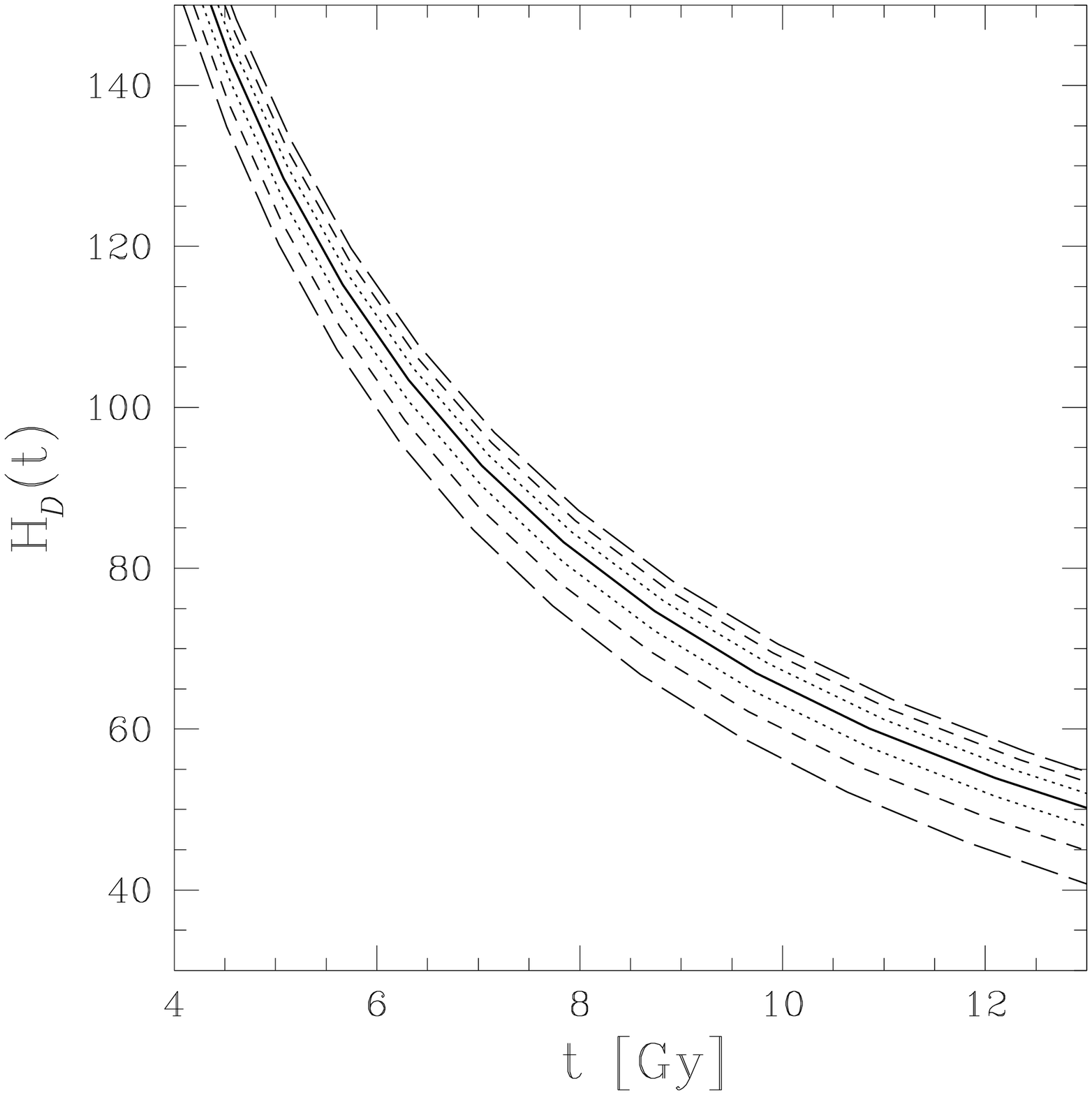,width=8cm}
\end{center}
\caption{\label{fig:HD-100Mpc}    The   evolution    of    the   
  Hubble--parameter $H_\CD(t)$  for a  spherical domain with  a scaled
  radius of  100Mpc.  Over--dense domains show a  smaller $H_\CD$ than
  in  an Einstein--de--Sitter universe  (solid line),  under--dense an
  increased     $H_\CD$.     The     same     conventions    as     in
  Fig.~\ref{fig:aD-zel-50Mpc} are used.}
\end{figure}

From  Fig.~\ref{fig:HD-100Mpc}   we  see   that  at  present   a  20\%
variability  of the  Hubble--parameter $H_\CD$  is quite  common  in a
domain  with  a  scaled  radius  of  100Mpc  (three--$\sigma$  initial
conditions).  For  a domain with a  scaled radius of  250Mpc still 3\%
variation is possible.  This agrees  well with the results obtained by
Wu et  al.~\cite{wu:measurement} for a  regional low--density universe
(see also {}\cite{nakamura:probability}, {}\cite{turner:relation}, and
refs.\ therein).

\begin{figure}
\begin{center}
\epsfig{figure=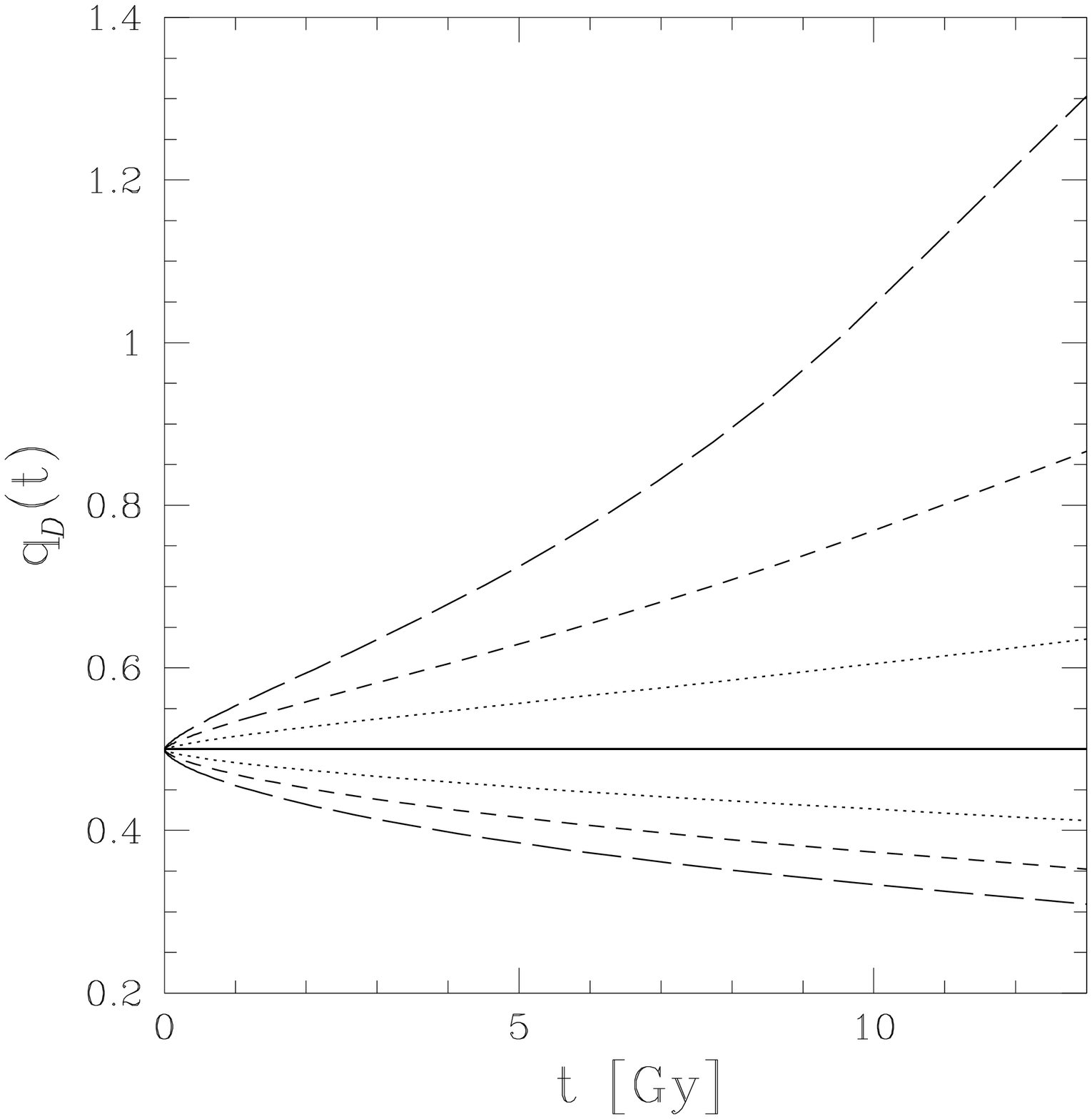,width=8cm}
\end{center}
\caption{\label{fig:qD-100Mpc} The evolution of the deceleration
  parameter $q_\CD(t)$ for a spherical  domain with a scaled radius of
  100Mpc.  Over--dense  domains show an increased  $q_\CD$ compared to
  an Einstein--de--Sitter universe  (solid line), under--dense domains
  a    decreased    $q_\CD$.     The    same   conventions    as    in
  Fig.~\ref{fig:aD-zel-50Mpc} apply.}
\end{figure}

The effect  becomes more pronounced if one  considers the deceleration
parameter $q_\CD$, as shown  in Fig.~\ref{fig:qD-100Mpc}.  In a domain
with a  scaled radius  of 100Mpc  the $q_\CD$ may  show a  present day
value   of   more   than   200\%   different   from   the   background
Einstein--de--Sitter universe.  Even for a domain with a scaled radius
of 250Mpc a difference larger than 15\% is possible.

\begin{figure}
\begin{center}
\epsfig{figure=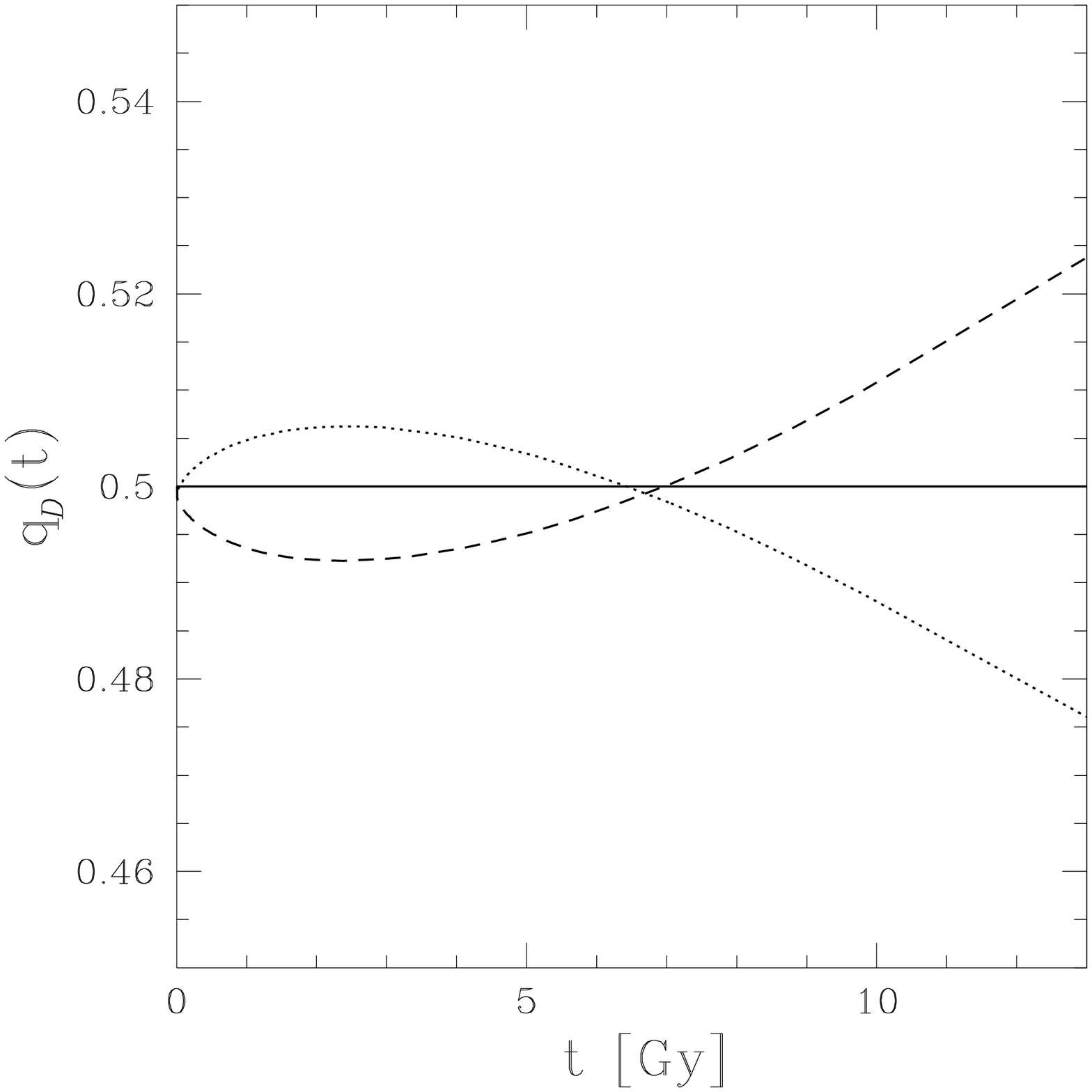,width=8cm}
\end{center}
\caption{\label{fig:qD-strange} The evolution of the regional deceleration
  parameter $q_\CD(t)$ for a spherical  domain with a scaled radius of
  100Mpc.   The initial  conditions  are chosen  as 0.5--$\sigma$  for
  $\baverage{\initial{\inI}}$    and    three--$\sigma$    for    both
  $\baverage{\initial{\inII}}$  and $\baverage{\initial{\inIII}}$ with
  the  opposite sign  of  $\baverage{\initial{\inI}}$.  The  initially
  accelerating,           i.e.\           under--dense          region
  ($\baverage{\initial{\inI}}>0$)  finally  decelerates (dotted  line)
  whereas  the initially  over--dense  region (dashed  line) shows  an
  accelerated expansion at late times.}
\end{figure}

Consider  the   situation  that  our  regional  Universe   may  be  an
under--dense  portion  (say,   resulting  from  an  under--density  at
0.5--$\sigma$ on the  scale of the considered portion).   Then, we may
well  live   in  a  shear--dominated   part  of  the   Universe:  many
low--amplitude anisotropic structures surround us which are compatible
with a regional under--density, but would suggest a dominance by shear
fluctuations (say,  resulting from a positive second  invariant in the
initial   conditions   at   three--$\sigma$).    Such   domains   with
density--contrast and  velocity fluctuations working  ``against'' each
other may show a qualitative  change in their evolution as illustrated
in   Fig.~\ref{fig:qD-strange}:  an  initially   under--dense  region,
however, dominated by shear fluctuations may show a deceleration.

\subsection{Evolution of the density parameters}
\label{sect:evolution-density-par}

Now  we turn  our  attention to  the domain--dependent  ``cosmological
parameters'' defined in Subsect.~\ref{sect:qualitative-discussion}.
Knowing the time evolution of  $a_\CD$ for given initial conditions we
can  calculate the time  evolution of  $\Omega_m^\CD$, $\Omega_k^\CD$,
and $\Omega_Q^\CD$.  Again we assume that the global evolution follows
an             Einstein--de--Sitter             model             with
$\Omega_\Lambda=0=\Omega_\Lambda^\CD$   and  on   the   largest  scale
$\Omega_k=0$.  However,  as shown below, for finite  domains $\CD$ the
time  evolution of the  ``cosmological parameters''  can substantially
differ  from  the  Friedmann   evolution  even  for  small,  seemingly
negligible   backreaction  term.   We   have  for   the  density   and
``curvature'' parameter:
\begin{align}
\Omega_m^\CD(t) = & \frac{H^2(\initial{t})\ (1-\laverage{\initial{\inI}})}
{{\dot a}_\CD(t)^2 a_\CD(t)} ,
\nonumber \\
\Omega_k^\CD(t) = & -\frac{k_\CD}{{\dot a}_\CD(t)^2} ,
\end{align}
with  $k_\CD=(\Omega_m^\CD(\initial{t})-1){\dot a}_\CD(\initial{t})^2$.
The evolution of $\Omega_Q^\CD$ is given by Eq.~\eqref{eq:def-omegaQ}.

For a homogeneous and  isotropic distribution of matter $\Omega_k^\CD$
may be related to the total  energy inside the domain $\CD$ within the
Newtonian interpretation.   This is not  possible in the  more general
inhomogeneous setting  we considered.   Now $k_\CD$ is  an integration
constant given by the initial conditions.  The role of $k_\CD$ and its
relation to  the average curvature in the  relativistic framework will
be explained in Sect.~\ref{sect:GR}.

\begin{figure}
\begin{center}
\epsfig{figure=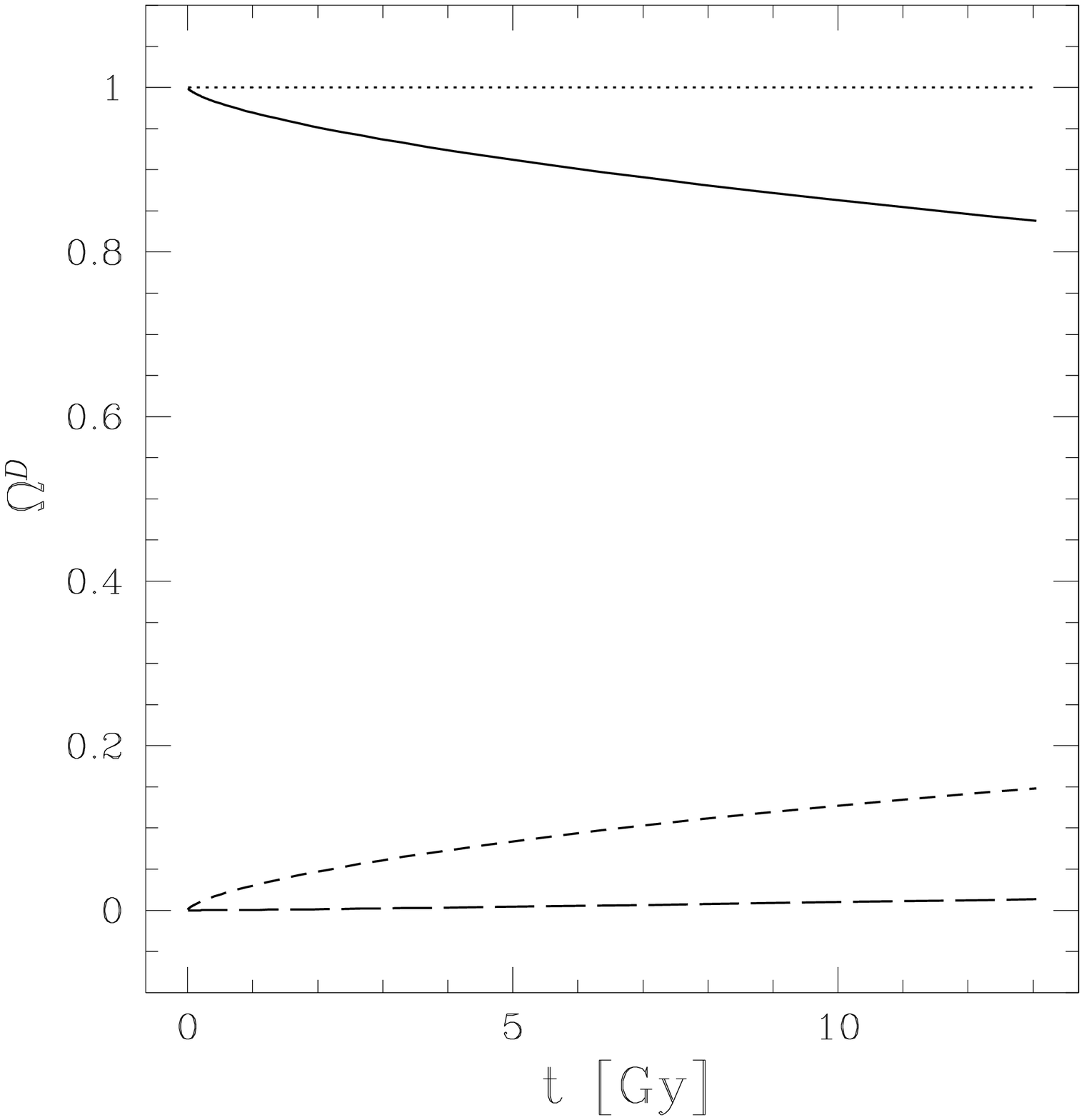,width=4.2cm}
\epsfig{figure=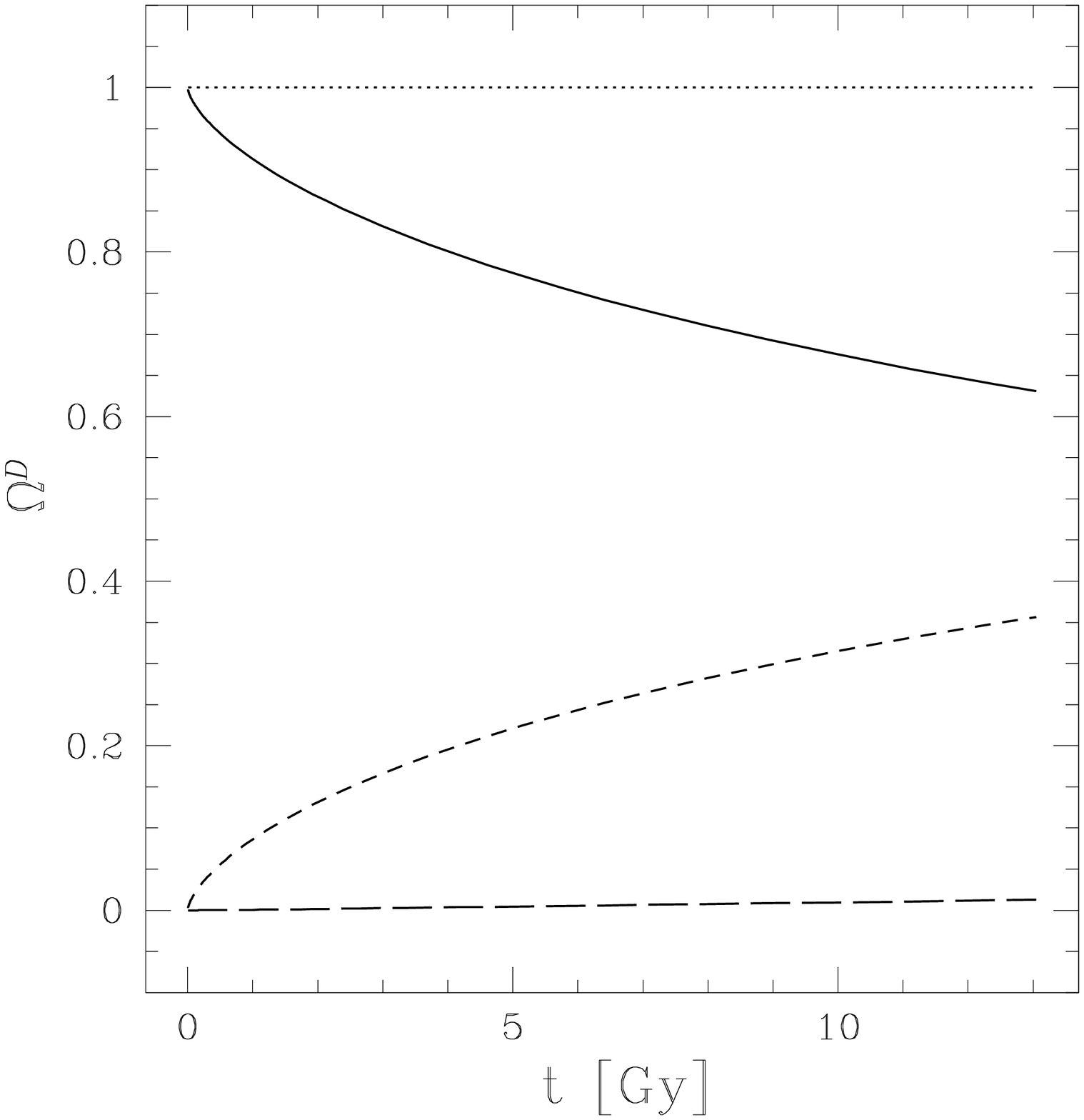,width=4.2cm}
\end{center}
\caption{\label{fig:omegas-expand} The evolution of the ``cosmological
  parameters''  $\Omega_m^\CD$  (solid  line),  $\Omega_k^\CD$  (short
  dashed),   and  $\Omega_Q^\CD$   (long  dashed)   in   an  expanding
  (under--dense) domain with initial radius 0.5Mpc (a scaled radius of
  100Mpc).         The       dotted       line        is       marking
  $\Omega_m^\CD+\Omega_k^\CD+\Omega_Q^\CD$.
  The  left  plot  is  for  one--$\sigma$,  the  right  plot  is  for
  three--$\sigma$  fluctuations.  The  corresponding evolution  of the
  scale factors is shown in Fig.~\ref{fig:aD-zel-100-250Mpc}.}
\end{figure}

\begin{figure}
\begin{center}
\epsfig{figure=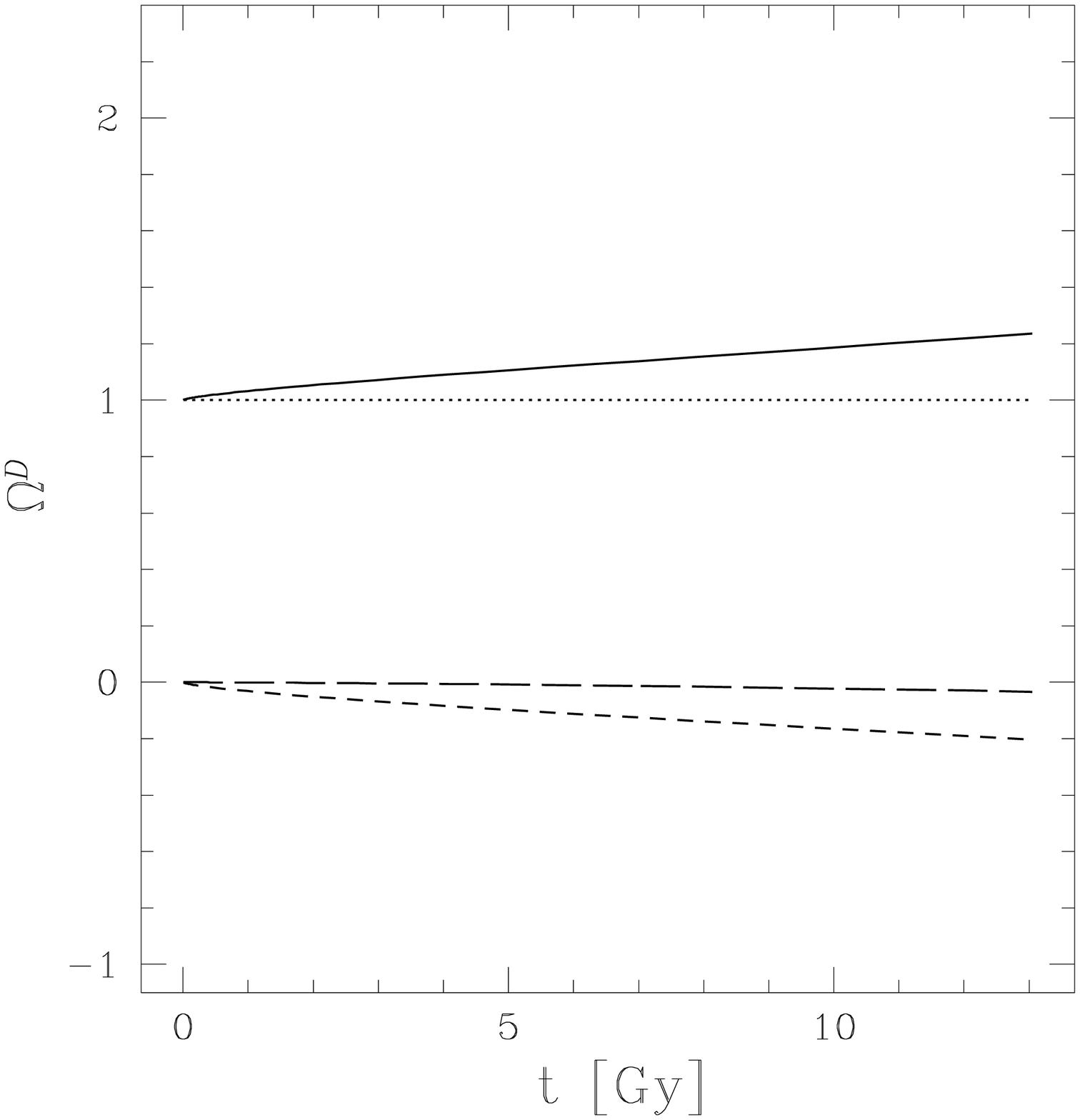,width=4.2cm}
\epsfig{figure=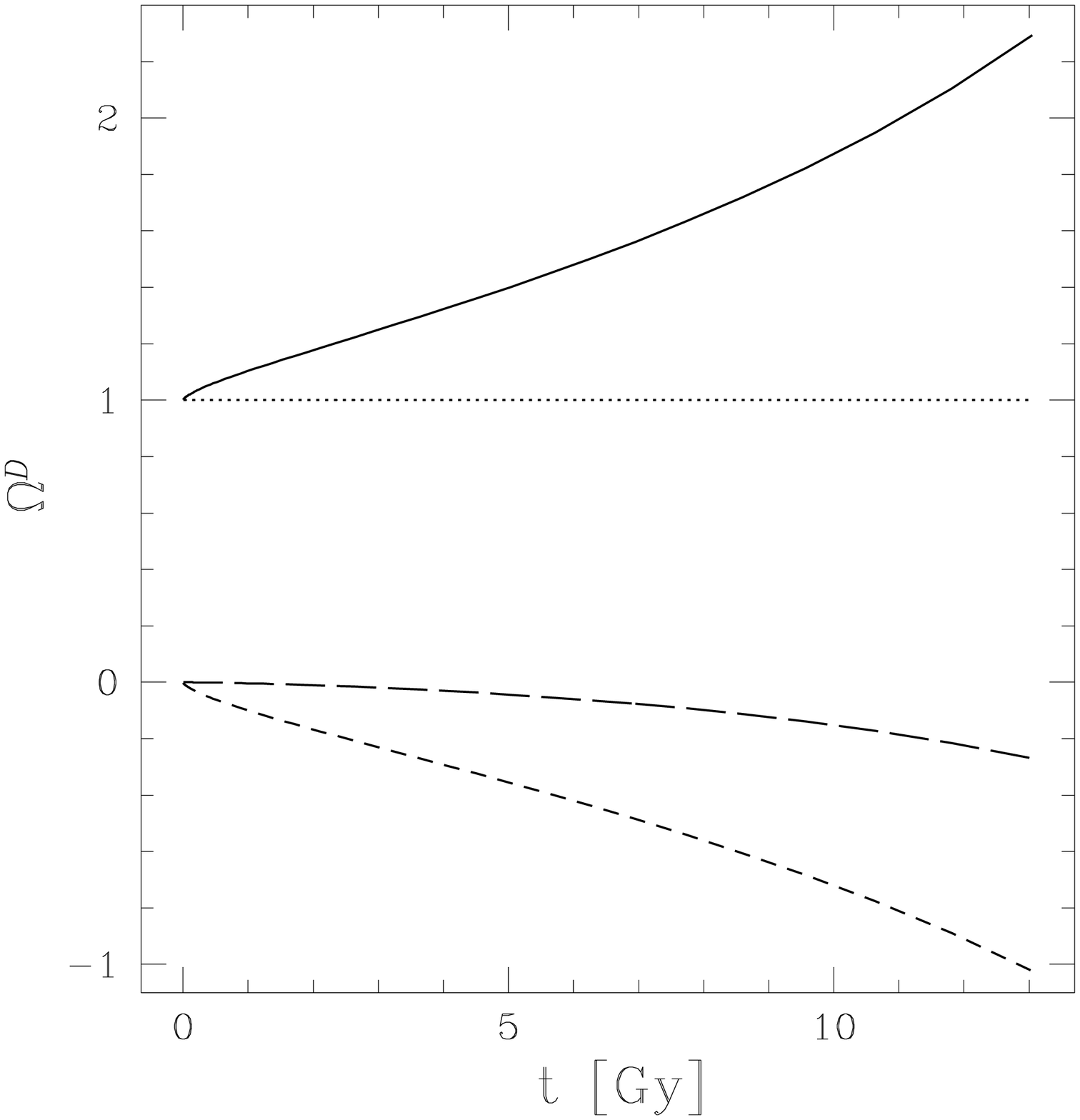,width=4.2cm}
\end{center}
\caption{\label{fig:omegas-collapse}   The  same   quantities   as  in
Fig.~\ref{fig:omegas-expand} are shown, now for a collapsing domain.}
\end{figure}

A  remarkable result is  that for  an accelerating  i.e.\ under--dense
region,    $\Omega_Q^\CD$    may   be    negligible    as   seen    in
Fig.~\ref{fig:omegas-expand},  but  dramatic   changes  in  the  other
parameters are observed. For a  collapsing domain with a scaled radius
of 100Mpc  the mass parameter  of the domain  may even differ  by more
than      100\%      from      the     global      mass      parameter
(Fig.~\ref{fig:omegas-collapse}).

\subsection{Self--consistency of the approximations}

In the preceding subsections we  calculated the evolution of the scale
factor    $a_\CD$   by    integrating   the    generalized   Friedmann
equation~\eqref{eq:expansion-law}   for  a  given   backreaction  term
$Q_\CD$.  We used exact  backreaction terms for the spherical collapse
and  the plane collapse,  but also  the approximate  backreaction term
$Q_\CD^Z$ for generic initial conditions.  This last approach must not
lead   to  consistent  results.    In  the   following  we   test  for
self--consistency    in     the    spirit    of     Doroshkevich    et
al.~\cite{doroshkevich:nonlinear} and  show that our  results obtained
with the ``Zel'dovich approximation'' are reliable.

Using $Q_\CD^Z$ and  integrating Eq.~\eqref{eq:evolution-ad} we obtain
an approximate solution $a_\CD^Z$ for the scale factor.
There is also the ``passive'' volume based approximation for the scale
factor     in      Zel'dovich's     approximation     according     to
Eq.~\eqref{eq:averageJZ}.  This kinematical $a_\CD^{\rm kin}$ does not
describe the  same evolution as  $a_\CD^Z$.  Both solutions  are exact
for       the       plane       collapse      initial       conditions
$\initial{\inII}=0=\initial{\inIII}$.   Moreover, $a_\CD^Z$ reproduces
the          (exact)          spherical         collapse          (see
Subsect.~\ref{sect:zeldovich-spherical-exact}),       whereas      the
kinematical solution $a_\CD^{\rm kin}$  does not.  Hence, we typically
rely on $a_\CD^Z$.

The  normalized  difference  
\begin{equation}
\epsilon_1 = \frac{|a_\CD^Z-a_\CD^{\rm  kin}|}{a_\CD^Z}
\end{equation}
may  serve as  an  internal consistency  check.   For initial  domains
larger  than  0.08Mpc (a  domain  with  scaled  radius of  16Mpc)  the
$\epsilon_1$ stayed well below 0.1  at all times, as long as $a_\CD^Z$
did not approach zero.

If $a_\CD^Z$ differs significantly  from the Friedmann solution $a$ we
are also interested in the impact of the inconsistency between the two
solutions  $a_\CD^Z$  and $a_\CD^{\rm  kin}$  on  the actual  observed
deviation:
\begin{equation}
\epsilon_2 = \frac{|a_\CD^Z-a_\CD^{\rm  kin}|}{|a_\CD^Z-a|} .
\end{equation}
For  all situations  with  $|a_\CD^Z-a|\gg0$, and  especially for  the
cases  considered   in  Fig.~\ref{fig:aD-zel-50Mpc},  $\epsilon_2$  is
always smaller than 0.15.

Hence,  the  results presented  above  are self--consistently  correct
within the Zel'dovich approximation.  Moreover, we can reproduce exact
solutions for both the spherical and the plane collapse, i.e.\ for two
orthogonal symmetry requirements.

\section{Some remarks}
\label{sect:some-remarks}

\subsection{Clues from general relativity}
\label{sect:GR}

The advent  of the ``backreaction  problem'' has its roots  in general
relativity   {}\cite{ellis:relativistic}.     Certainly,   a   general
relativistic  treatment has to  be considered  when talking  about the
global  effect  of  backreaction.   We  have shown  that  a  Newtonian
treatment falls short of explaining the global effect.  We remark that
the same applies for a general relativistic treatment that is designed
as  close to  the Newtonian  one as  possible.  For  example,  Russ et
al.~\cite{russ:age}  have  worked   with  perturbation  schemes  on  a
3--Ricci  flat background  employing periodic  boundary  conditions on
some large scale in order  to estimate the global backreaction effect. 
In  their  investigation  the backreaction  term  (Eq.~\eqref{eq:Q-GR}
below)  vanishes by  construction  on the  periodicity  scale and  its
global effect cannot  be estimated (see {}\cite{buchert:onaverage} for
further discussion).   The question whether an  averaged space section
can be  identified at all with  a Fried\-mann--Lema\^\i{}tre cosmology
on some large  scale, even approximately, is open  and can probably be
answered  only on the  basis of  a full  background--free relativistic
investigation.

Although it  is of  particular importance to  find strategies  for the
estimation of this global effect,  the impact of backreaction on small
scales is also interesting from  a general relativistic point of view. 
We  will discuss  hints in  this subsection  that show  that,  even on
scales that  are previously thought to be  satisfactorily described by
Newton's theory,  general relativity can,  and we think, will  play an
exciting  role.   On  top   of  previous  estimations  of  the  global
backreaction effect in GR we have to acknowledge that the aspect added
by   Buchert  and   Ehlers  {}\cite{buchert:averaging},   namely  that
backreaction   terms   appear   naturally   also  by   averaging   out
inhomogeneous matter distributions in Newtonian theory, is paraphrased
in the  general relativistic framework  as well. We  shall concentrate
below on this correspondence.

Let us first briefly review  the corresponding equations that rule the
averaged  dynamics in  general relativity  {}\cite{buchert:onaverage}. 
Given a foliation of  spacetime into flow--orthogonal hypersurfaces we
can derive equations analogous  to the generalized Friedmann equations
discussed in this paper (restricting attention to irrotational flows).
Spatial averaging of any scalar  field $\Psi$ is a covariant operation
given a foliation of spacetime and is defined as follows:
\begin{equation}
\label{eq:average-GR}
\average{\Psi (t, X^i)}: = 
\frac{1}{V_\CD}\int_\CD J d^3 X \;\;\Psi (t, X^i) \;\;\;,
\end{equation}
with  $J:=\sqrt{\det(g_{ij})}$, where  $g_{ij}$ is  the metric  of the
spatial  hypersurfaces, and  $X^i$ are  coordinates that  are constant
along  flow lines.   The spatially  averaged equations  for  the scale
factor $a_\CD$, respecting mass conservation, read:\\
averaged Raychaudhuri equation:
\begin{equation}
\label{eq:expansion-law-GR}
3\frac{{\ddot a}_\CD}{a_\CD} + 4\pi G 
\frac{M_\CD}{\initial{V}a_\CD^3} - \Lambda = {Q}_\CD;
\end{equation}
averaged Hamiltonian constraint:
\begin{equation}
\label{eq:hamiltonconstraint}
\left( \frac{{\dot a}_\CD}{a_\CD}\right)^2 - \frac{8\pi G}{3}
\frac{M_\CD}{\initial{V}a_\CD^3} + \frac{\average{\CR}}{6} 
- \frac{\Lambda}{3} = -\frac{Q_\CD}{6} ,
\end{equation}
where   the  mass   $M_\CD$,   the  averaged   spatial  Ricci   scalar
$\average{\CR}$   and   the    ``backreaction   term''   $Q_\CD$   are
domain--depen\-dent and, except the mass, time--depen\-dent functions.
The backreaction source term is given by
\begin{equation}
\label{eq:Q-GR} 
Q_\CD : = 2 \average{\inII} - \frac{2}{3}\average{\inI}^2 =
\frac{2}{3}\average{\left(\theta - \average{\theta}\right)^2 } - 
2\average{\sigma^2} .
\end{equation}
Here,  $\inI$  and $\inII$  denote  the  invariants  of the  extrinsic
curvature  tensor that  correspond  to the  kinematical invariants  we
employed.  The  same  expression  (except  for the  vorticity)  as  in
Eq.~\eqref{eq:Q-kinematical-scalars} follows  by introducing the split
of the  extrinsic curvature into  the kinematical variables  shear and
expansion (second equality above).

We  appreciate an  intimate correspondence  of the  GR  equations with
their   Newtonian   counterparts   (Eq.~\eqref{eq:expansion-law}   and
Eq.~\eqref{eq:average-friedmann}).  The  first  equation  is  formally
identical  to  the  Newtonian   one,  while  the  second  delivers  an
additional   relation   between  the   averaged   curvature  and   the
backreaction  term that  has no  Newtonian analogue.  This  implies an
important  difference   that  becomes  manifest  by   looking  at  the
time--derivative     of     Eq.~\eqref{eq:hamiltonconstraint}.     The
integrability  condition   that  this  time--derivative   agrees  with
Eq.~\eqref{eq:expansion-law-GR}  is nontrivial in  the GR  context and
reads:
\begin{equation}
\label{eq:integrability-GR}
\partial_t Q_\CD + 6 \frac{{\dot a}_\CD}{a_\CD} Q_\CD +  
\partial_t \average{\CR}
+ 2 \frac{{\dot a}_\CD}{a_\CD} \average{\CR} = 0 .
\end{equation}

The correspondence  between the Newtonian  $k_\CD$--pa\-ra\-meter with
the averaged 3--Ricci curvature is  more involved in the presence of a
backreaction term:
\begin{equation}
\label{eq:integrability-integral-GR}
\frac{k_\CD}{a_\CD^2} - \frac{1}{3 a_\CD^2} \int_{t_0}^t \,dt' \;
Q_\CD\; \frac{d}{dt'} a^2_\CD(t')
= \frac{1}{6}\left(\langle {\cal R} \rangle_\CD + Q_\CD\right) . 
\end{equation}

The  time--derivative  of Eq.~\eqref{eq:integrability-integral-GR}  is
equivalent        to         the        integrability        condition
Eq.~\eqref{eq:integrability-GR}.        Eq.~\eqref{eq:integrability-GR}
shows  that  averaged curvature  and  backreaction  term are  directly
coupled  unlike in  the  Newtonian case,  where the  domain--dependent
$k_\CD$--parameter is  fixed by the initial  conditions. For initially
vanishing $k_\CD$ in Newton's theory, the ``curvature parameter'' will
always  stay zero.  This is  not  the case  in the  GR context,  where
backreaction produces  averaged curvature  in the coarse  of structure
formation,  even in  the  case of  domains  that are  on average  flat
initially.  In  view of our results that  the $k_\CD$--parameter could
play  an important  role to  compensate for  under--densities  in some
domain in  a globally flat universe,  the GR context  suggests that we
shall have  a more complex  situation with regard to  the backreaction
term itself.   It remains to  be seen whether this  coupling increases
the quantitative relevance of the backreaction term itself, and how it
affects the dynamics of spatial  domains also on ``Newtonian scales''. 
This will be the subject of a forthcoming work.

\subsection{N--body simulations}
\label{sect:n-body}

Most cosmological N--body simulations solve the Newtonian equations of
motion in comoving coordinates  typically inside a cuboid region $\CC$
with periodic  boundary conditions.  Hence,  the boundary of  $\CC$ is
empty, $\partial\CC=\emptyset$,  and from Eq.~\eqref{eq:Q-surface-int}
we directly obtain $Q_\CC=0$.
For  the definition  of  comoving coordinates  a background  expansion
factor  $a(t)$  is  required.   Consistent  with  $Q_\CC=0$  the  time
evolution of $a_\CC(t)$  is assumed to follow the  time evolution of a
Friedmann--Lema\^\i{}tre background  model $a(t)$ throughout  the {\em
  whole}  simulation  process.  Hence,  N--body  simulations give  the
correct results (within their  resolution limits) for the evolution of
inhomogeneities in  domains where no  backreaction is present  at {\em
  all} times.
As  we  have seen  in  the  preceding  sections, for  generic  initial
conditions  the backreaction  may  influence the  global expansion  of
domains even with a present size of hundreds of Mpc's.  The difference
in  the time--evolution  of $a_\CD$  compared to  $a$ acts  as  a time
dependent source for the peculiar--potential $\phi$:
\begin{equation}
\triangle_q \phi = 4\pi G a^2 \varrho_H \delta \ + \ 
3\frac{a}{a_\CD}\left({\ddot a}_\CD a -  a_\CD {\ddot a}\right) .
\end{equation}
A  general  domain  will  also  change its  shape  as  illustrated  in
Fig.~\ref{fig:domains}.   This  will  also  change  the  evolution  of
inhomogeneities inside  these volumes. Therefore,  N--body simulations
only trace a limited subset of the full space of solutions.
Moreover,   the   fluctuations   in   the   initial   conditions   are
underestimated   for    periodic   boundaries   as    illustrated   in
Sect.~\ref{sect:sigma-numbers}.

It will be a challenge  to incorporate backreaction effects in N--body
simulations.  In  order to improve the estimation  of the backreaction
effect, one  could embed  high--resolution N--body simulations  into a
model for  the large scales such  as the model  investigated here.  In
this line  Takada and Futamase {}\cite{takada:local}  have suggested a
hybrid  model  that  incorporates  the full  nonlinearities  on  small
scales, while the  large scales are described by  perturbation theory. 
A  similar procedure  was  already used  for high--resolution  cluster
simulations,   surrounded  by   inhomogeneities   determined  from   a
low--resolution cosmological simulation {}\cite{bartelmann:arcII}.
However,  abandoning the  Fast--Fourier--Transform  with its  periodic
boundaries, as typically used for solving the Poisson equation in PM--
and P$^3$M--simulations, will be only one technical obstacle.  In this
respect     tree--codes    are     more     flexible    (see     e.g.\ 
{}\cite{hernquist:ewald} for the treatment of periodic boundaries).

Another question is, whether we are able to estimate the effect of the
backreaction    using    simulations.     Consider   a    partitioning
$\CD_n\subset\CC$  of   the  periodic  box  $\CC$   into  $N$  subsets
intersecting only at their boundaries with $\CC=\bigcup_{n=1}^N\CD_n$.
Typically these $\CD_n$ are sub--cubes of the periodic box.  We obtain
for  the  full   backreaction  from  Eqs.~\eqref{eq:Q-surface-int}  or
{}\eqref{eq:backreaction-comoving}:
\begin{equation}
\label{eq:nbody-cancel}
|\CC| Q_\CC = \sum_{n=1}^N |\CD_n| Q_{\CD_n} = 0.
\end{equation}
In  general  each $Q_{\CD_n}\ne0$,  but  they  cancel  over the  whole
simulation box. Let us assume that all the $\CD_n$ have the same shape
and        volume        $|\CD_n|=|\CC|/N$.        According        to
Eq.~\eqref{eq:nbody-cancel}  the mean  backreaction  for such  domains
estimated by
\begin{equation}
\frac{1}{N}\sum_{n=1}^N Q_{\CD_n}
\end{equation}
will always  be zero  in simulations. An  estimate of the  variance is
possible by
\begin{equation}
\label{eq:nbody-var}
\frac{1}{N}\sum_{n=1}^N Q_{\CD_n}^2 \ne 0 .
\end{equation}
However, the backreaction terms $Q_{\CD_n}$ of the domains $\CD_n$ are
correlated        through       Eq.~\eqref{eq:nbody-cancel},       and
Eq.~\eqref{eq:nbody-var}   underestimates    the   variance   of   the
backreaction  especially for  domains  $\CD_n$ comparable  in size  to
$\CC$.

\section{Discussion, Conclusions and Outlook}

We presented a quantitative  investigation of the backreaction effect,
i.e.,  the   deviations  of   dynamical  properties  of   an  averaged
inhomogeneous  model from  that of  a  homogeneous--isotropic standard
cosmology.   We did  this  for scalar  observables  on finite  spatial
domains such  as the scale  factor and its derivatives  (the expansion
and deceleration  rates), as well  as for derived parameters  such as
the  cosmological  density  parameters   defined  on  the  domains  of
averaging.

Although  we did  this for  generic initial  conditions common  in the
modeling  of large--scale structure,  our approach  is limited  by the
Newtonian approximation,  the setting of  periodic boundary conditions
on the inhomogeneities  on some large scale, and  the restriction to a
first--order  Lagrangian  perturbation  scheme  for  the  modeling  of
inhomogeneities.   We may  consider  the proposed  approximation as  a
prototype  model of  any averaged  inhomogeneous cosmology  asking for
refinements in the restrictions mentioned.

First of all we learned that the dynamics of a domain is controlled by
two effective sources for the expansion rate governed by a generalized
Friedmann  equation: the first  is the  over--/under--density compared
with the global density,  usually studied by the spherically symmetric
top--hat  model.   This  effect  can  be  absorbed  into  the  initial
conditions  and the  average model  remains  within the  class of  the
standard  Friedmann--Lema\^\i{}tre  solutions.    The  second  is  the
backreaction term itself that measures the departure from the class of
standard models.  It consists of fluctuations in the kinematical parts
of the  velocity gradient, the expansion  rate and the  rates of shear
and  vorticity.  
Consistently,  the spherical  model with  an over--  or under--density
inside the  spherical domain  describes the isotropic  deviations from
the Friedmann--Lema\^\i{}tre evolution,  as expressed by the vanishing
backreaction term.  This interpretation  is valid if one considers the
evolution  of the whole  spherical domain.   Locally however,  we have
contributions from the expansion rate $\theta$ and the shear $\sigma$,
canceling only  on average in  the spherical domain.   Additionally to
these  isotropic deviations  the plane--collapse  or the  more general
models     allow    for     anisotropic     deviations    from     the
Friedmann--Lema\^\i{}tre evolution.  One  may express this directly in
terms  of  the  averaged  invariants: $\average{\inI}=3H_\CD$  is  the
isotropic deviation from  the Hubble expansion $3H$; $\average{\inII}$
incorporates  the shear responsible  for anisotropic  deviations. Both
the   fluctuations  in   the  expansion   rate   $\average{\inI}$  and
$\average{\inII}$ contribute to the backreaction term.

On  the basis of  a first--order  Lagrangian perturbation  approach we
proposed an average model for generic initial conditions that is exact
in  two orthogonal  cases,  the plane  and  the spherically  symmetric
evolution.  According  to these  properties the new  model generalizes
the spherical top--hat model.  Comparisons showed that the collapse is
typically accelerated in this more  general setting.  At least down to
the limit scale of the  evolution model our results should be correct,
i.e.  for truncated high--frequency  end in the initial power spectrum
this     is     roughly    the     scale     of    galaxy     clusters
{}\cite{weiss:optimizing}.

By  evaluating   the  expectation  values  for   the  fluctuations  we
quantified the  expected magnitudes of  the parts of  the backreaction
term given  a power spectrum  of initial density fluctuations  (in our
case we have chosen a standard CDM power spectrum).  The effect may be
viewed from  two perspectives: a one--$\sigma$ fluctuation  of a given
part of  the backreaction on  some smaller scale may  be alternatively
viewed  as a  two--$\sigma$  fluctuation of  a correspondingly  larger
domain.  E.g., a negative one--$\sigma$  fluctuation in all parts on a
domain   of  a  current   radius  of   $100$Mpc  (with   a  normalized
Hubble--parameter of $0.5$, a global  density parameter of unity and a
zero global $k-$parameter) will  give today a matter density parameter
of $0.84$ compensated by a $k-$parameter of roughly $0.15$.  We see in
this example that the  backreaction parameter itself can be negligible
quantitatively  ($\Omega_Q^\CD=0.01$),  but  qualitatively  it  has  a
strong impact  on the  change of the  other parameters.   We emphasize
that this  is a large  effect, since a  $15\%$ change of  the standard
parameters on a corresponding cubic  volume of about $160$Mpc is a lot
for the ``innocent'' expectation values chosen at one--$\sigma$.

These  initial fluctuations  driving the  non--standard  evolution may
arise from  modes of  the density fluctuations  inside or  outside the
domain, their  relative strength depending  on the size of  the domain
and  the  shape of  the  power--spectrum  considered.   Using the  CDM
power--spectrum on  small scales, where  modes larger than  the domain
with a radius of 100Mpc are set to zero, the fluctuations of e.g.\ the
first  invariant  are  reduced  by  a  factor of  two  to  three  (see
Fig.~\ref{fig:sigmas}).   Clearly the  large--scale  fluctuations give
the  major contribution,  but also  fluctuations inside  contribute at
least 30\%  in this case.  Hence,  the feasibility of  a split between
small--  and  large--scale  contributions  critically depends  on  the
spectrum and the size of the volume considered.

Consider, e.g., the MarkIII  and SFI catalogs of peculiar--velocities. 
Adopting  our Hubble constant  these samples  roughly correspond  to a
radial depth  only slightly larger  than $100$Mpc.  According  to what
has been  said before  we may likely  live in an  ``untypical'' region
when talking about  $100$Mpc.  (The likelihood of such  events has not
been  quantified in the  present paper,  but is  a subject  of ongoing
work).   We therefore  think that  matching results  from small--scale
data   with,    e.g.,   the   supernovae    constraints   (see   e.g.\ 
{}\cite{zehavi:evidence}) has  to be premature  unless we do  not have
either  a considerably larger  sample, or  else, know  the fluctuation
properties of the regional Universe by other observations.
Since  the value  of  the  backreaction term  $Q_\CD$  depends on  the
velocity  field inside  $\CD$,  these peculiar--velocity catalogs  may
offer  the  possibility  of estimating $Q_\CD$. 
In this line two papers are of particular interest:
by taking  the sampling anisotropies of the  velocity field explicitly
into  account,  Reg\"os and  Szalay  {}\cite{regos:multipole} found  a
large effect  ($40\%$) of the  dipol and quadrupol anisotropies  on the
estimated   bulk   flow   of   the   elliptical   galaxy   sample   of
{}\cite{davies:spectroscopy};
using     the     Eulerian     linear     approximation     G{\'o}rski
{}\cite{gorski:pattern}   showed   that    the   velocity   field   is
significantly correlated even on scales of $100$Mpc.

Furthermore, we  may well  deal with an  underestimate of  the effect,
since our  assumptions are  conservative in a  number of  respects:\\
The ``Zel'dovich approximation'' reliably traces large--scale features
but  lacks  structure  on small  scales  which  would  add up  in  the
averages,  since  shear fluctuations  and  expansion fluctuations  are
positive definite.   Similar remarks apply to the  small--scale end of
the  power spectrum,  which  differs between  models.  Certainly,  the
spectral  index on  large scales  and the  position of  the bend--over
plays a crucial role.   Thus, as a first point we may  say that in the
family of models  usually studied the standard CDM  model evolved with
the ``Zel'dovich approximation'' may be regarded as conservative in
these respects.\\
We  started  our calculations  at  $\initial{z}=200$  since at  higher
redshifts the  (linear) power--spectrum is  still changing its  shape. 
Therefore, we have to assume  that the evolution of the domain follows
the Friedmann  equation with no backreaction up  to $\initial{z}=200$. 
Our matter model  (dust) is not a good  description for earlier times,
however   extrapolating  and   starting  at   $\initial{z}=1000$,  the
resulting deviations increase only slightly, well below the percent
range for the examples considered.\\
As a third point, a conservative constraint that we have put in is the
existence of a  Hubble flow on some large  scale.  From our discussion
of  the general  relativistic case  it is  suggested that  a realistic
inhomogeneous cosmology will  take its freedom to drift  away from the
standard model.  Assuming $Q_\CD=0$ for  all times means that we force
the effect  to add up to  zero within the  whole model for all  times. 
This certainly hints towards an  increase of the effect under study in
more realistic
situations.\\
Fourth, due  to the Newtonian investigation we  were also conservative:
in general relativity, the  curvature parameter is directly coupled to
the backreaction parameter.  Since  we have learned that the Newtonian
$k_\CD$--parameter plays a crucial role, this remark means that we may
underestimate  the  curvature  effect   (this  has  to  be  confirmed,
however).\\
Fifth, we have been talking about ``typical'' fluctuations, e.g., when
we specified  a one--$\sigma$ amplitude of  a given peculiar--velocity
gradient invariant.   Especially regional observational  data reaching
up to $100$Mpc could well be those of an untypical region lying in the
tail of  the ({\em  non--Gaussian}) probability distributions  for the
(second and third) invariants.

In this line we would like to point out that, even if the fluctuations
in number  density (the first  moment of the galaxy  distribution) may
not  be present, fluctuations  may show  up in  higher moments  of the
galaxy distribution.  A recent  investigation of subsets from the IRAS
1.2~Jy   catalog  revealed   large  fluctuations   in   the  Minkowski
functionals    as     well    as    the     two--point    correlations
{}\cite{kerscher:fluctuations}    at   least    up   to    scales   of
200$h^{-1}$Mpc.  The  IRAS 1.2~Jy  catalog is an  example of  a sample
that may look ``homogeneous'' as far as the number density of galaxies
is  concerned  {}\cite{davis:homogeneous},  but features  fluctuations
already  for the  second moment  and,  most dramatic,  for the  higher
moments.

A number of open questions are to be considered.  We already mentioned
the quantification of the likelihood to  find a region of a given size
for  given  expectation values  of  the  invariants.  The  probability
distributions of the averaged initial invariants would allow us to set
likelihood constraints on the cosmological parameters.
Another  related task  is  to  quantify cosmic  variance:  due to  the
scale--dependence  of  the   ``cosmological  parameters''  we  can  in
principle infer their variations  as a function of scale.  Controlling
these  variations allows  a statistical  assessment of  how and  if we
approach a  ``scale of homogeneity''. Results  from spatial variations
of parameters in an averaged  inhomogeneous model may then be compared
with observed fluctuations in galaxy or cluster catalogs.

We  have  seen  that  the   backreaction  term,  although  it  may  be
numerically   small,   significantly   modifies   the   evolution   of
cosmological parameters,  driven by  the fluctuations in  the velocity
field.   We  expect  exciting   new  insights  by  moving  to  general
relativistic average models, a subject of forthcoming work.

\section*{Acknowledgements}

We  would like  to thank  Claus  Beisbart and  Toshifumi Futamase  for
interesting and helpful discussions.  TB acknowledges generous support
and hospitality by the  National Astronomical Observatory in Tokyo, as
well  as  hospitality  at  Tohoku  University in  Sendai,  Japan.   MK
acknowledges  support from  the  {\em Sonderforschungsbereich  SFB 375
f{\"u}r Astroteilchenphysik}  and CS from the  grant {\em MU~1536/1-1}
of DFG.


\appendix
\section{Derivation of the generalized Friedmann equation of Newtonian 
cosmology}
\label{sect:generalized-friedmann}

In  this  appendix we  give  a  short  derivation of  the  generalized
Friedmann equation {}\cite{buchert:averaging}.

In  Eulerian  space  one  of  the dynamical  fields  is  the  velocity
$\bv(\bx,t)$.   It is  suitable  to introduce  the  rate of  expansion
$\theta =  \nabla \cdot \bv$, the shear  $(\sigma_{ij})$, and rotation
${\vec \omega} = \frac{1}{2} \nabla\times\bv$ via a decomposition of
the velocity gradient\footnote{A comma denotes partial derivative with
  respect  to  Eulerian  coordinates $\partial/\partial  x_i\equiv,i$;
  $\delta_{ij}$  is  the  Kronecker  delta  and  $\epsilon_{ijk}$  the
  totally   antisymmetric  unit   tensor;  we   adopt   the  summation
  convention.}:
\begin{equation}
v_{i,j} = \sigma_{ij} + \tfrac{1}{3} \delta_{ij} \theta + \omega_{ij} ;
\quad \omega_{ij} = -\epsilon_{ijk} \omega_{k} ; 
\quad \sigma_{[ij]} = 0 .
\end{equation}
The  rate of  shear $\sigma$  and the  rate of  rotation  $\omega$ are
defined as follows:
\begin{equation}
\sigma^2 = \tfrac{1}{2}\ \sigma_{ij} \sigma_{ij}\ , \
\omega^2 = \tfrac{1}{2}\ \omega_{ij} \omega_{ij} .
\end{equation}

Now,  consider the volume  of an  Eulerian spatial  domain $\CD$  at a
given  time, $V=\int_{\CD}\rmd^3x$,  and follow  the  position vectors
$\bx=\mbf(\bX,t)$ of  all fluid  elements (labelled by  the Lagrangian
coordinates  $\bX$  within  the  domain). Then,  each  volume  element
changes according to $\rmd^3x=J\rmd^3X$,  where $J$ is the determinant
of the  transformation from  Eulerian to Lagrangian  coordinates.  The
total rate  of change of  the volume of  the same collection  of fluid
elements of a domain may then be calculated as follows:
\begin{multline}
\label{eq:theta-HD}
\frac{d_t V}{V} = \frac{1}{V} \rmd_t \int_{\initial{\CD}}\rmd^3X\ J 
= \frac{1}{V}\int_{\initial{\CD}}\rmd^3X\  d_t J \\ 
= \frac{1}{V} \int_{\initial{\CD}}\rmd^3X\ \theta J 
= \average{\theta} = 3 H_{\CD},
\end{multline}
where   $d_t$   is  the   total   (Lagrangian)  time--derivative   and
$H_{\CD}={\dot  a}_\CD/a_\CD$  the  Hubble--parameter  of  the  domain
$\CD$.

It  is crucial  to notice  that  the total  time--derivative does  not
commute with spatial averaging. For an arbitrary tensor field $\CA$ we
derive  with   the  help  of  the  above   definitions  the  following
commutation rule:
\begin{equation}
\label{eq:commutation-rule}
\rmd_{t}\average{\CA}-\average{\rmd_{t}\CA}=
\average{\CA\theta}-\average{\theta}\average{\CA}.
\end{equation}
The generalized Friedmann equation Eq.~\eqref{eq:expansion-law} of the
main     text     together      with     the     backreaction     term
Eq.~\eqref{eq:Q-kinematical-scalars}  follows by  setting $\CA=\theta$
and  using   the  mass  conservation   Eq.~\eqref{eq:average-rho}  and
Raychaudhuri's equation for the  local evolution of the expansion rate
$\theta$:
\begin{equation}
\label{eq:raychaudhuri}
\dot\theta=\Lambda-4\pi G\rho -\frac{1}{3}\theta^2 + 2 (\omega^2 - 
\sigma^2) .
\end{equation}
(Eq.~\eqref{eq:raychaudhuri}  follows from  the trace  of  the spatial
derivative  of  Euler's   equation  $\rmd_t\bv=\bg$  using  the  field
equation $\nabla\cdot\bg=\Lambda-4\pi  G\varrho$ for the gravitational
field strength $\bg$.)

\section{Invariants}
\label{sect:invariants}

The   three   principal   scalar   invariants  of   a   tensor   field
$\CA=(A_{ij})$ in Cartesian coordinates are defined by
\begin{align}
\inI(A_{ij})   & = {\rm tr}(A_{ij}) , \nonumber \\
\inII(A_{ij})  & = \tfrac{1}{2} 
  \left( {\rm tr}(A_{ij})^2 - {\rm tr}( (A_{ij})^2 ) \right) , \nonumber \\
\inIII(A_{ij}) & = {\rm det}(A_{ij}) \nonumber .
\end{align}
In particular  for $A_{ij}=v_{i,j}$ the invariants  are expressible in
terms of kinematical  scalars and can be written  as total divergences
of vector fields, which has been  used and discussed in the context of
perturbation                                                  solutions
{}\cite{buchert:lagrangianthree,ehlers:newtonian}:
\begin{align}
\label{eq:v-inv-I}
\inI(v_{i,j}) & = v_{i,i} = \nabla \cdot \bv = \theta ,\\[1ex]
\label{eq:v-inv-II}
\inII(v_{i,j})
& = \tfrac{1}{2}\left((v_{i,i})^2 - v_{i,j}v_{j,i}\right) \nonumber \\
& = \tfrac{1}{2}\nabla\cdot
\Big(\bv (\nabla\cdot\bv) - (\bv\cdot\nabla)\bv \Big)\\
& = \omega^2 - \sigma^2 + \tfrac{1}{3} \theta^2 , \nonumber \\[1ex]
\label{eq:v-inv-III}
\inIII(v_{i,j}) 
& = \tfrac{1}{3} v_{i,j}v_{j,k}v_{k,i} - \tfrac{1}{2}(v_{i,i})
(v_{i,j}v_{j,i})+\tfrac{1}{6}(v_{i,i})^3 \nonumber \\
& = \tfrac{1}{3}\nabla\cdot\left( \tfrac{1}{2}\nabla\cdot
\Big( \bv(\nabla\cdot\bv) - (\bv\cdot\nabla)\bv \Big) \bv - \right.\nonumber \\
&\qquad \qquad \left.
\Big(\bv(\nabla\cdot\bv) - (\bv\cdot\nabla)\bv \Big)\cdot\nabla\bv \right) \\
& = \tfrac{1}{9}\theta^3 + 2\theta \left(\sigma^2 + 
\tfrac{1}{3}\omega^2 \right)
+ \sigma_{ij}\sigma_{jk}\sigma_{ki} 
- \sigma_{ij}\omega_i \omega_j . \nonumber
\end{align}
These expressions are also valid for comoving quantities
\begin{equation}
\inI(\partial_{q_j}u_i) = \nabla_q \cdot \bu\quad  \text{ etc. },
\end{equation}
with              the              comoving              differentials
$\partial_{q_j}=\frac{\partial}{\partial q_j}$ and $\nabla_q$.

\section{Initial conditions}
\label{sect:initial}

To estimate  the magnitude of the volume--averages  of the invariants,
we    first    relate    the    initial    peculiar--velocity    field
$\bU(\bX)=\bu(\bq,\initial{t})$   to  the  initial   density  contrast
$\initial{\delta}(\bX)=(\varrho(\bX,\initial{t})-
{}\varrho_H(\initial{t}))/\varrho_H(\initial{t})$  using  the Eulerian
linear approximation.

At time  $\initial{t}$ the  comoving coordinates equal  the Lagrangian
coordinates  $\bq=\bX$ by  definition.  Considering  only  the growing
mode        in        an        Einstein--de--Sitter        background
$\delta(\bq,t)=(t/\initial{t})^{2/3}\initial{\delta}(\bq)$,         the
peculiar--expansion       rate        evolves       according       to
$\theta_u(\bq,t)=(t/\initial{t})^{1/3}\initial{\theta}(\bq)$ with
\begin{equation}
\initial{\theta}(\bq) = \frac{-2}{3\ \initial{t}} \initial{\delta}(\bq) .
\end{equation}
Assuming a non--rotational initial velocity field
\begin{align}
\label{eq:initial-u-delta}
\bu(\bq,\initial{t}) 
& = \int_{\BR^3}\rmd^3k\ 
\initial{\tilde\theta}(\bk)\frac{\bk}{ik^2}\rme^{i\bk\cdot\bq} \nonumber\\
& = -\frac{2}{3\ \initial{t}} \int_{\BR^3}\rmd^3k\ 
\initial{\tilde\delta}(\bk)\frac{\bk}{ik^2}\rme^{i\bk\cdot\bq},
\end{align}
with $k=|\bk|$  and the Fourier transform  $\widetilde{\CA}(\bk)$ of a
field $\CA(\bq)$:
\begin{align}
\widetilde{\CA}(\bk) =& \frac{1}{(2\pi)^3}\ \int_{\BR^3}\rmd^3q\ 
\CA(\bq)\rme^{-i\bk\cdot\bq} .
\end{align}
In  an  Einstein--de--Sitter  background  the  non--rotational  initial
velocity  field  is  related  to  the initial  displacement  field  by
$\psi_{|ij}=3/2\         \initial{t}U_{i|j}$,         and         with
Eq.~\eqref{eq:initial-u-delta} we  get the invariants  of $\psi_{|ij}$
in terms of the initial density contrast $\initial{\delta}$.

\subsection{Volume--averaged invariants}
\label{sect:vol-averaged-inv}

To simplify  calculations we  consider spherical domains  $\CB_R$ with
radius $R$.   Since we  assume that the  perturbations in  the initial
density  field  are   stochastically  independent  from  the  specific
location in space, we may center  the sphere on the origin.  We define
the normalized window function (a top--hat)
\begin{equation}
W_R(\bX) = \left\{
\begin{array}{ll}
1/|\CB_R| & \text{if } \bX\in\CB_R,\\
0 & \text{else},
\end{array} 
\right.
\end{equation}
where $|\CB_R|=4\pi/3\ R^3$ is the volume of  the sphere. The Fourier 
transform of $W_R$,
\begin{equation}
\widetilde{W}_R(k) = \frac{2}{|\CB_R|\ (2\pi)^2\ k^3}
\left(\sin(kR)-kR\cos(kR)\right),
\end{equation}
is    depending   on    $k=|\bk|$   only.     The    spatial   average
$\baverage{\CA(\bX)}$  of   a  field  $\CA$   may  be  written   as  a
convolution:
\begin{align}
\baverage{\CA} =& \int_{\BR^3}\rmd^3X\ \CA(\bX)W_R(\bX) \nonumber \\
=& (2\pi)^3 \int_{\BR^3}\rmd^3k\ \widetilde{\CA}(\bk) 
\widetilde{W}_R(k) .
\end{align}

With                                    Eq.~\eqref{eq:initial-u-delta},
{}\eqref{eq:v-inv-I}--{}\eqref{eq:v-inv-III},                       and
{}\eqref{eq:def-initial-inv} we can express the averaged invariants in
terms  of the  initial density  contrast  $\initial{\delta}$ (remember
$\bU(\bX)=\bu(\bq,\initial{t})$          and          $\psi_{|ij}=3/2\ 
\initial{t}U_{i|j}$):
\begin{equation}
\label{eq:initial-I-delta}
\baverage{\initial{\inI}} = -(2\pi)^3\
\int_{\BR^3}\rmd^3k\ \widetilde{\initial{\delta}}(\bk) \widetilde{W}_R(k) ,
\end{equation}
\begin{multline}
\label{eq:initial-II-delta}
\baverage{\initial{\inII}} = \frac{(2\pi)^3}{2}\
\int_{\BR^3}\rmd^3k_1 \int_{\BR^3}\rmd^3k_2\ \\
\widetilde{\initial{\delta}}(\bk_1)
\widetilde{\initial{\delta}}(\bk_2)\ \widetilde{W}_R(|\bk_1+\bk_2|)
\left(1- \frac{(\bk_1\cdot\bk_2)^2}{k_1^2 k_2^2}\right) ,
\end{multline}
and 
\begin{multline}
\label{eq:initial-III-delta}
\baverage{\initial{\inIII}} = -(2\pi)^3\ 
\int_{\BR^3}\rmd^3k_1 \int_{\BR^3}\rmd^3k_2 \int_{\BR^3}\rmd^3k_3 \\
\widetilde{\initial{\delta}}(\bk_1) 
\widetilde{\initial{\delta}}(\bk_2) 
\widetilde{\initial{\delta}}(\bk_3)\ 
\widetilde{W}_R(|\bk_1+\bk_2+\bk_3|) F(\bk_1,\bk_2,\bk_3),
\end{multline}
with
\begin{multline}
F(\bk_1,\bk_2,\bk_3) = 
\frac{1}{6} - \frac{1}{2}\frac{\left(\bk_2\cdot\bk_3\right)^2}{k_2^2 k_3^2} +\\
+ \frac{1}{3}\frac{(\bk_1\cdot\bk_2)(\bk_1\cdot\bk_3)(\bk_2\cdot\bk_3)}
{k_1^2 k_2^2 k_3^2}.
\end{multline}
Similar expressions without the  spatial averaging were presented in a
systematic  way by {}\cite{goroff:coupling,makino:analytic}  (see also
{}\cite{jain:second}).

To   calculate    the   evolution   of   the    scale   factor   using
Eq.~\eqref{eq:evolution-ad} we also  need the normalized over--density
$\average{\delta}=(\average{\varrho}-\varrho_H)/\varrho_H$.        With
Eq.~\eqref{eq:average-rho}                                          and
$\varrho_H(t)=\varrho_H(\initial{t})/a(t)^3$     we     can    express
$\average{\delta}$ in terms of the averaged first invariant:
\begin{equation}
\label{eq:Delta-I}
\average{\delta}
= \frac{a^3}{a_\CD^3}\left(1+\laverage{\delta(\initial{t})}\right) -1 
= \frac{a^3}{a_\CD^3}\left(1-\laverage{\initial{\inI}}\right) -1  .
\end{equation}

\subsection{Distribution of the volume--averaged invariants}

In  Appendix~\ref{sect:vol-averaged-inv}  we  expressed  the  averaged
invariants  of the peculiar--velocity  field as  volume--averages over
the initial density contrast $\initial{\delta}$.
To  get  a  handle  on  the  distribution  of  these  volume--averaged
invariants we  calculate their ensemble mean and  their variance under
the assumption that $\initial{\delta}$ is a Gaussian random field.
A  Gaussian   density  field   is  stochastically  specified   by  its
power-spectrum    $\initial{P}(k)$   with    $k=|\bk|$    (see   e.g.\ 
{}\cite{sahni:approximation}),  determining  the  correlation  of  the
Fourier modes of the initial density contrast:
\begin{equation}
\label{eq:powerspectrum}
\BE[\initial{\widetilde{\delta}}(\bk)\initial{\widetilde{\delta}}(\bk')]
= \delta^D(\bk+\bk')\ \initial{P}(k) .
\end{equation}
$\delta^D(\cdot)$ is  the Dirac delta--distribution.   Contrary to the
{\em spatial  average} $\baverage{\cdot}$ over a  {\em finite} domain,
$\BE[\cdot]$ is denoting the {\em ensemble average}.  Since a Gaussian
field is  ergodic {}\cite{adler:randomfields}, a  spatial average over
the  whole  space  $\langle\cdot\rangle_{\BR^3}$ equals  the  ensemble
average  $\BE[\cdot]$.  However,  we  focus on  spatial averages  over
compact spherical domains $\CB_R$.

By the  definition of a Gaussian  random field all odd  moments of the
density  contrast vanish.  I.e.\  $\BE[\initial{\delta}(\bX)^n]=0$ for
$n$                   odd,                   and                  from
Eqs.~(\ref{eq:initial-I-delta}--\ref{eq:initial-III-delta}) follows
\begin{equation}
\label{eq:odd-moments}
\begin{array}{lcl}
\BE[\baverage{\initial{\inI}}] & = 
0 = & \BE[\baverage{\initial{\inI}} \baverage{\initial{\inII}}] ,\\[1ex]
\BE[\baverage{\initial{\inIII}}] & =
0 = & \BE[\baverage{\initial{\inII}} \baverage{\initial{\inIII}}] . 
\end{array}
\end{equation}
This   already   tells   us   that   $\baverage{\initial{\inI}}$   and
$\baverage{\initial{\inII}}$, as  well as $\baverage{\initial{\inII}}$
and  $\baverage{\initial{\inIII}}$   are  uncorrelated,  and   may  be
specified independently.

The other invariants and their products require some calculations:\\
With  Eq.~\eqref{eq:initial-I-delta},  {}\eqref{eq:powerspectrum}, and
{}\eqref{eq:odd-moments} we obtain
\begin{align}
\label{eq:initial-I-Pk}
\sigma_{\inI}^2(R) & = 
\BE\left[\left(
\baverage{\initial{\inI}}-\BE[\baverage{\initial{\inI}}]
\right)^2\right] 
= \BE[\baverage{\initial{\inI}}^2] \nonumber \\
& = (2\pi)^{6}
\int_{\BR^3}\rmd^3k\ \initial{P}(k)\widetilde{W}_R(k)^2 .
\end{align}
The  variance $\sigma_{\inI}^2(R)$  is equal  to the  well--known mean
square  fluctuations  of  the  {\em initial}  density  contrast  field
smoothed with a top--hat of radius $R$.\\
For the second invariant we obtain
\begin{multline}
\BE[\baverage{\initial{\inII}}] = 
2 (2\pi)^3 \int_{\BR^3}\rmd^3k_1\ \initial{P}(k)\times \\
\int_{\BR^3}\rmd^3k_2\ \delta^D(\bk_2) 
\left(1- \frac{(\bk_1\cdot\bk_2 - k_1^2)^2}{k_1^2 (\bk_2-\bk_1)^2}\right)
\widetilde{W}_R(k_2) .
\end{multline}
Using spherical coordinates and  taking the limit $k_2\rightarrow0$ in
the last integral, we arrive at
\begin{equation}
\BE[\baverage{\initial{\inII}}] = 0.
\end{equation}
For  a Gaussian  density  field the  four--point correlation  function
factorizes into two--point correlations as expressed by
\begin{multline}
\label{eq:fourpoint-fact}
\BE[\initial{\widetilde{\delta}}(\bk_1)\initial{\widetilde{\delta}}(\bk_2)
\initial{\widetilde{\delta}}(\bk_3)\initial{\widetilde{\delta}}(\bk_4)] =\\
\begin{array}{rl}
\initial{P}(k_1)\initial{P}(k_3)& 
\delta^D(\bk_1+\bk_2)\delta^D(\bk_3+\bk_4) \\
+ \initial{P}(k_1)\initial{P}(k_2)&
\delta^D(\bk_1+\bk_3)\delta^D(\bk_2+\bk_4) \\
+ \initial{P}(k_1)\initial{P}(k_2)&
\delta^D(\bk_1+\bk_4)\delta^D(\bk_2+\bk_3).
\end{array}
\end{multline}
Then,  the variance  of the  second invariant  can be  calculated from
Eq.~\eqref{eq:initial-II-delta}:
\begin{multline}
\label{eq:initial-II-Pk}
\sigma_{\inII}^2(R)=\BE[\baverage{\initial{\inII}}^2] = \\
\frac{(2\pi)^6}{2} \int_{\BR^3}\rmd^3k_1 \int_{\BR^3}\rmd^3k_2\
\initial{P}(k_1)\initial{P}(k_2)\\
\widetilde{W}_R(|\bk_1+\bk_2|)^2
\left(1- \frac{(\bk_1\cdot\bk_2)^2}{k_1^2 k_2^2}\right)^2 .
\end{multline}
A similar calculation gives 
\begin{equation}
\BE[\baverage{\initial{\inI}} \baverage{\initial{\inIII}}] = 0,
\end{equation}
showing          that          $\baverage{\initial{\inI}}$         and
$\baverage{\initial{\inIII}}$  are  uncorrelated,  and may  be  chosen
independently for a Gaussian random field.\\
Similar  to   Eq.~\eqref{eq:fourpoint-fact}  one  may   factorize  the
six--point function for  a Gaussian density field. This  enables us to
express  $\BE[\baverage{\initial{\inIII}}^2]$ in  terms  of the  power
spectrum. After some lengthy algebra we obtain
\begin{multline}
\label{eq:initial-III-Pk}
\sigma_{\inIII}^2(R)=\BE[\baverage{\initial{\inIII}}^2] = \\
(2\pi)^6 \int_{\BR^3}\rmd^3k_1 \int_{\BR^3}\rmd^3k_2 \int_{\BR^3}\rmd^3k_3\\
\initial{P}(k_1)\initial{P}(k_2)\initial{P}(k_3)\
\Big[\widetilde{W}_R(k_1)^2 G_1(\bk_1,\bk_2,\bk_3) +\\ 
\widetilde{W}_R(|\bk_1+\bk_2+\bk_3|)^2 G_2(\bk_1,\bk_2,\bk_3)\Big],
\end{multline}
with
\begin{multline}
G_1(\bk_1,\bk_2,\bk_3) = \\
-2 \frac{(\bk_1\cdot\bk_2)^2(\bk_1\cdot\bk_3)^2}{k_1^4k_2^2k_3^2}
+\frac{13}{4}\frac{(\bk_1\cdot\bk_2)^2}{k_1^2k_2^2}
-\frac{5}{4}, 
\end{multline}
and 
\begin{multline}
G_2(\bk_1,\bk_2,\bk_3) = \\
\frac{2}{3} \frac{(\bk_1\cdot\bk_2)^2(\bk_1\cdot\bk_3)^2(\bk_2\cdot\bk_3)^2}
{k_1^4k_2^4k_3^4}\\
-2 \frac{(\bk_1\cdot\bk_2)^3(\bk_1\cdot\bk_3)(\bk_2\cdot\bk_3)}
{k_1^4k_2^4k_3^2}\\
+ \frac{2}{3} \frac{(\bk_1\cdot\bk_2)(\bk_1\cdot\bk_3)(\bk_2\cdot\bk_3)}
{k_1^2k_2^2k_3^2}\\
+\frac{2}{3} \frac{(\bk_1\cdot\bk_2)^4}{k_1^4k_2^4}
- \frac{(\bk_1\cdot\bk_2)^2}{k_1^2k_2^2}
+\frac{1}{6} .
\end{multline}

\subsection{Some Numbers}
\label{sect:sigma-numbers}

To  calculate the  fluctuations  $\sigma_{\inI}(R)$ etc.\  we need  to
specify the  initial power spectrum $\initial{P}(k)$.   Models for the
linearly evolved  power spectrum of the  density contrast fluctuations
at present time $t_0$ are typically given by
\begin{equation}
\label{eq:power-transfer}
P(k, t_0) = A k^n\ T(k, t_0)^2 ,
\end{equation}
with  the amplitude  $A$, the  primordial spectral  index $n$  and the
transfer function $T(k, t_0)$.
We  use $n=1$  and  a transfer  function  for Cold  Dark Matter  (CDM)
parametrized  according to  Bardeen  et al.~\cite{bardeen:gauss},  for
dark  matter with  $\Omega_m=1$, a  small baryon  fraction,  and three
relativistic neutrino flavors:
\begin{multline}
T(k, t_0) = \frac{\ln(1+r k)}{r k}\times \\
\left(1 + s k + (t k)^2 + (u k)^3 + (v k)^4\right)^{-\frac{1}{4}} ,
\end{multline}
with   $r=9.36$Mpc,  $s=15.56$Mpc,   $t=64.4$Mpc,   $u=21.84$Mpc,  and
$v=26.84$Mpc.
Consistent        with       Sect.~\ref{sect:evolution-cosmo}       an
Einstein--de--Sitter  background  with  $H_0=50{\rm  km  s}^{-1}  {\rm
  Mpc}^{-1}$ is  assumed.  $k$  is in units  of ${\rm  Mpc}^{-1}$.  We
choose  the normalization  $A= 2.19\  10^4{\rm Mpc}^{4}$  resulting in
$\sigma(8/0.5\  {\rm  Mpc})=1$  (see  Eq.~\eqref{eq:initial-I-Pk}  and
Table~\ref{table:sigmas}).   This normalization  is  smaller than  the
COBE--normalization {}\cite{bunn:fouryear}, but  still larger than the
cluster--normalization {}\cite{viana:cluster} for this CDM--model.

\begin{figure}
\begin{center}
\epsfig{figure=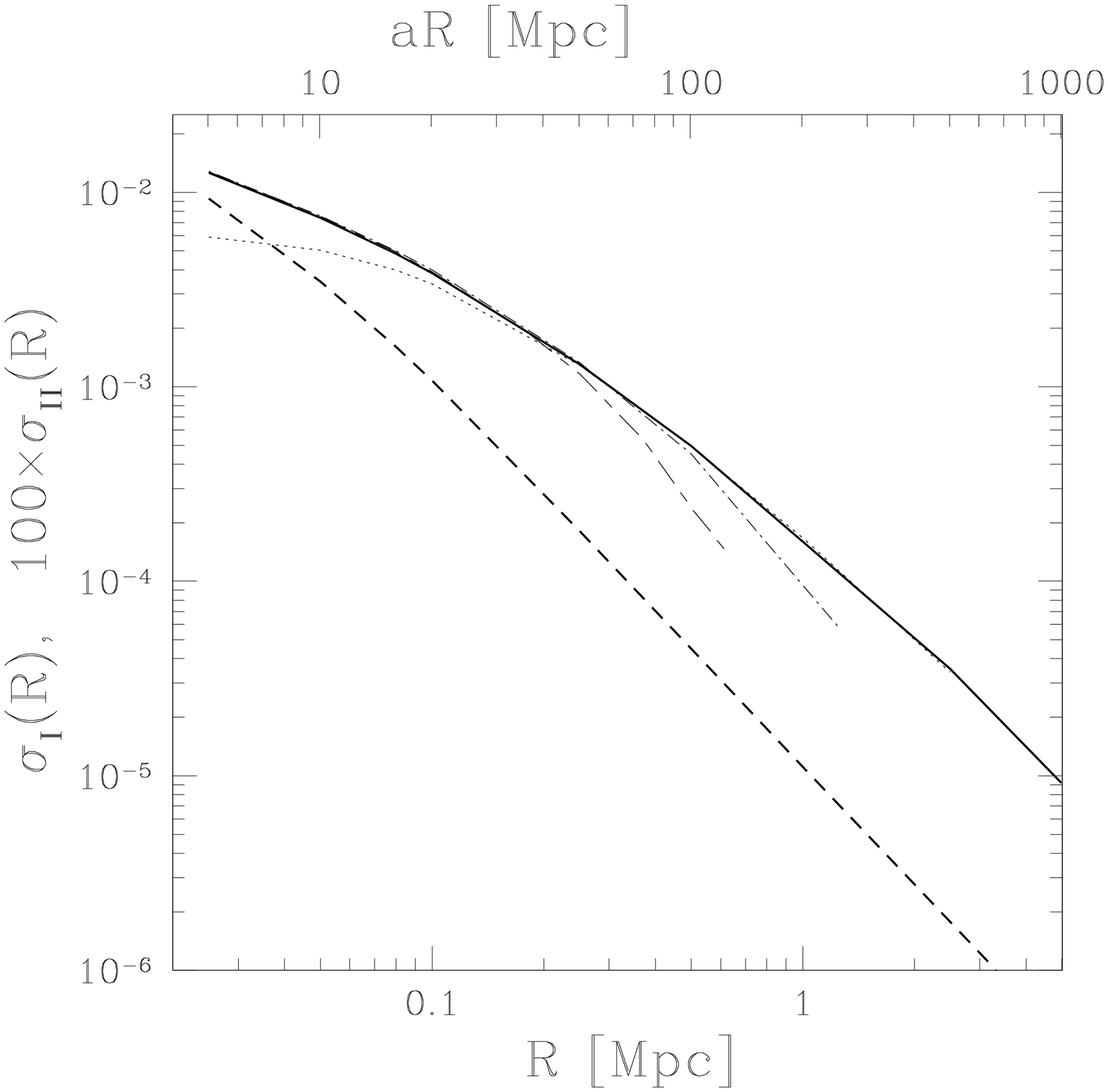,width=8cm}
\end{center}
\caption{ The values of $\sigma_{\inI}(R)$ and $\sigma_{\inII}(R)$ are
  shown  against  the  radius  $R$  of the  spherical  initial  domain
  $\CB_R$.  On  top also the  radius $a(t_0)R$, with  $a(t_0)=201$, of
  the linearly evolved domain is  given. The dotted line is the result
  for    $\sigma_{\inI}(R)$    using    the   truncated    power    of
  Eq.~\eqref{eq:truncatedP}.  The  short--dashed long--dashed line and
  the dashed--dotted  line correspond to  power spectra with a  cut at
  small $k$, corresponding to $L=200$Mpc and $L=400$Mpc respectively.}
\label{fig:sigmas}
\end{figure}

We choose our initial  conditions at $\initial{z}=200$, since the {\em
linear}  power spectrum does  not significantly  change its  shape for
$z<\initial{z}$   {}\cite{bond:collisionless}.    The   linear   power
spectrum  at initial time  $\initial{t}$, with  $a(\initial{t})=1$, is
directly   related   to   the   linear  power   spectrum,   given   in
Eq.~\eqref{eq:power-transfer} at present time $t_0$ by
\begin{equation}
\initial{P}(k) = \frac{P(k,t_0)}{a(t_0)^2}.
\end{equation}
Using this CDM--model we calculate $\sigma_{\inI}(R)$ and 
$\sigma_{\inII}(R)$ for initial domains of radius $R$ (see 
Fig.~\ref{fig:sigmas}).
Some  comments on  the  numerical methods  are  in order:  calculating
$\sigma_{\inI}(R)$  from Eq.~\eqref{eq:initial-I-Pk}  does  not impose
any  numerical problems.  We  used a  crude Monte--Carlo.   The triple
integral appearing in Eq.~\eqref{eq:initial-II-Pk} for the calculation
of $\sigma_{\inII}(R)$ required stratified sampling. In both cases our
results are accurate to one digit at least.
The        calculation       of        $\sigma_{\inIII}(R)$       from
Eq.~\eqref{eq:initial-III-Pk}  requires some  major  computational and
algorithmic efforts in order to evaluate the eight integrals.  This is
beyond the scope  of this article. We only need  an order of magnitude
estimate  for  $\sigma_{\inIII}(R)$. Guided  by  our  analysis of  the
spherical collapse model  (see Eq.~\eqref{eq:spherical-I-III}) we will
use  the estimate  $\sigma_{\inIII}(R)\approx1/27\ \sigma_{\inI}(R)^3$
in our calculations.
The   values    for   $\sigma_{\inI}(R)$,   $\sigma_{\inII}(R)$,   and
$\sigma_{\inIII}(R)$      used     in     our      calculations     in
Sect.~\ref{sect:quantifying} are given in Table~\ref{table:sigmas}.

It   has  been   shown   that  the   accuracy   of  the   ``Zel'dovich
approximation'' may  be increased by using a  smoothed initial density
field,     i.e.\     a      truncated     power     spectrum     $P_s$
{}\cite{coles:testing,melott:optimizing,buchert:testing,melott:testing,karakatsanis:temporal}.
The optimum smoothing scale for  CDM initial conditions is in our case
$k_s=1.687$Mpc$^{-1}$ {}\cite{weiss:optimizing}, resulting in
\begin{equation}
\label{eq:truncatedP}
P_s(k) = \rme^{-k^2/k_s^2}\ \initial{P}(k) .
\end{equation}
In Fig.~\ref{fig:sigmas}  we show $\sigma_{\inI}(R)$  calculated using
$P_s(k)$  instead of  $\initial{P}(k)$.   Already for  domains with  a
scaled radius larger than 20Mpc the difference becomes unimportant.

In a periodic box with sidelength $L$, the power spectrum is truncated
at  small $k$.   The suppression  of the  fluctuations in  the initial
conditions may be  estimated by using a cut:  $P(k)=0$ for $k<2\pi/L$. 
In  Fig.~\ref{fig:sigmas}   the  effect   of  applying  this   cut  to
$\sigma_{\inI}(R)$ is illustrated for scaled sidelengths $L=200$Mpc and
$L=400$Mpc,  corresponding  to  spherical  domains with  a  radius  of
$aR=125$Mpc and $aR=250$Mpc.

\end{document}